\documentclass[prl,superscriptaddress,nofootinbib,notitlepage,twocolumn,tightenlines]{revtex4-1}
\usepackage{xurl}
\PassOptionsToPackage{hyphens}{url}\usepackage[colorlinks=true,citecolor=blue,urlcolor=blue,linkcolor=blue,breaklinks]{hyperref}
\usepackage{graphicx}
\usepackage{amsmath}

\begin{document}
\title{Data-driven discovery and extrapolation of parameterized pattern-forming dynamics}
\author{Zachary G. Nicolaou}
\affiliation{Department of Applied Mathematics, University of Washington, Seattle, Washington 98195, USA}
\author{Guanyu Huo}
\affiliation{Department of Applied Mathematics, University of Washington, Seattle, Washington 98195, USA}
\author{Yihui Chen}
\affiliation{Department of Applied Mathematics, University of Washington, Seattle, Washington 98195, USA}
\author{Steven L. Brunton}
\affiliation{Department of Mechanical Engineering, University of Washington, Seattle, Washington 98195, USA}
\author{J. Nathan Kutz}
\affiliation{Department of Applied Mathematics, University of Washington, Seattle, Washington 98195, USA}

\begin{abstract}
Pattern-forming systems can exhibit a diverse array of complex behaviors as external parameters are varied, enabling a variety of useful functions in biological and engineered systems.  First-principles derivations of the underlying transitions can be characterized using bifurcation theory on model systems whose governing equations are known.  In contrast, data-driven methods for more complicated and realistic systems whose governing evolution dynamics are unknown have only recently been developed. Here we develop a data-driven approach,  the {\em sparse identification for nonlinear dynamics with control parameters} (SINDyCP), to discover dynamics for systems with adjustable control parameters, such as an external driving strength. We demonstrate the method on systems of varying complexity, ranging from discrete maps to systems of partial differential equations. To mitigate the impact of measurement noise, we also develop a weak formulation of SINDyCP and assess its performance on noisy data. We demonstrate applications including the discovery of universal pattern-formation equations, and their bifurcation dependencies, directly from data accessible from experiments and the extrapolation of predictions beyond the weakly nonlinear regime near the onset of an instability.\\[1.5em]
Phys. Rev. Research \textbf{5}, L042017 (2023) \\
DOI: \href{https://doi.org/10.1103/PhysRevResearch.5.L042017}{https://doi.org/10.1103/PhysRevResearch.5.L042017}

\end{abstract}

\maketitle

Data-driven approaches to system identification are undergoing a revolution, spurred by the increasing availability of computational resources, data, and the development of novel and reliable machine learning algorithms \cite{book_bk,2021_Karniadakis,2020_Udrescu}.
The {\em sparse identification of nonlinear dynamics} (SINDy) is a particularly simple and flexible mathematical approach that leverages efficient sparse optimization algorithms in the automated discovery of complex system dynamics and governing equations \cite{2016_Brunton}. 
In this Letter, we leverage the SINDy model discovery framework to understand parametric dependencies and underlying bifurcations in pattern-forming systems.  Specifically, we develop the SINDY with control parameters (SINDyCP) to discover such parameterized dynamics.

It has been 30 years since Cross and Hohenberg's seminal and authoritative review consolidating an exceptionally large body of work on pattern formation across a broad range of physical systems~ \cite{1993_Cross}. Universal equations determined by normal forms of canonical bifurcations \cite{1998_Kuznetsov}, such as the complex Ginzburg-Landau equation \cite{2002_Aranson}, govern the formation of patterns near the onset of instabilities across scientific disciplines. Such equations continue to reveal insights into complex systems, including in the study of, for example, synchronization, biophysics, active matter, and quantum dynamics  \cite{2017_Nicolaou, 2022_Heinonen}. 

Despite the success of pattern-formation theory in modeling complex dynamics, ongoing challenges remain in applying such model equations more broadly. First-principle derivations and the computation of normal-form parameters in terms of physical driving parameters are tedious, costly, and error prone. Furthermore, the resulting weakly nonlinear models are only theoretically justified near the onset of instability, while interesting and important pattern-forming processes often occur far from the instability threshold. Recent advances in data-driven system identification are opening new avenues of research to address these challenges, including a paradigm for modeling strongly nonlinear regimes beyond the asymptotic approximations reviewed by Cross and Hohenberg~\cite{1993_Cross}. 

The SINDy model discovery framework is particularly well suited to the modern analysis of bifurcations and normal forms, as it generates interpretable models that have as few terms as possible, balancing model complexity and descriptive capability.  
A variety of extensions of the SINDy approach have been developed since its introduction.  For example, SINDYc enables the discovery of systems subject to external control signals \cite{2016_Brunton2,2018_Kaiser,2021_Fasel}, while PDEFind \cite{2017_Rudy,2017_Schaeffer} enables the discovery of spatiotemporal dynamics characterized by partial differential equations (PDEs). Data-driven approaches can also learn to disambiguate between parametric dependency and governing equations and discover bifurcations~\cite{2016_Brunton,2017_Schaeffer_2, 2019_Rudy}. Model pattern-formation equations typically encode the effects of external drive through a number of driving parameters, which characterize the bifurcation leading to the onset of instability.  Several recent works establish system identification on pattern-forming systems ranging from closure models for fluid turbulence \cite{2020_Schmelzer, 2020_Zanna, 2021_Beetham,2017_Kramer} to biochemical reactions and active matter systems \cite{2020_Wang, 2021_Romeo, 2021_Supekar}. These approaches show promise, but \textit{crucially}, they have not demonstrated the ability to extrapolate by detecting pattern-forming instabilities that may develop when driving parameters differ from those used in the training data.  

\begin{figure*}
\includegraphics[width=2\columnwidth]{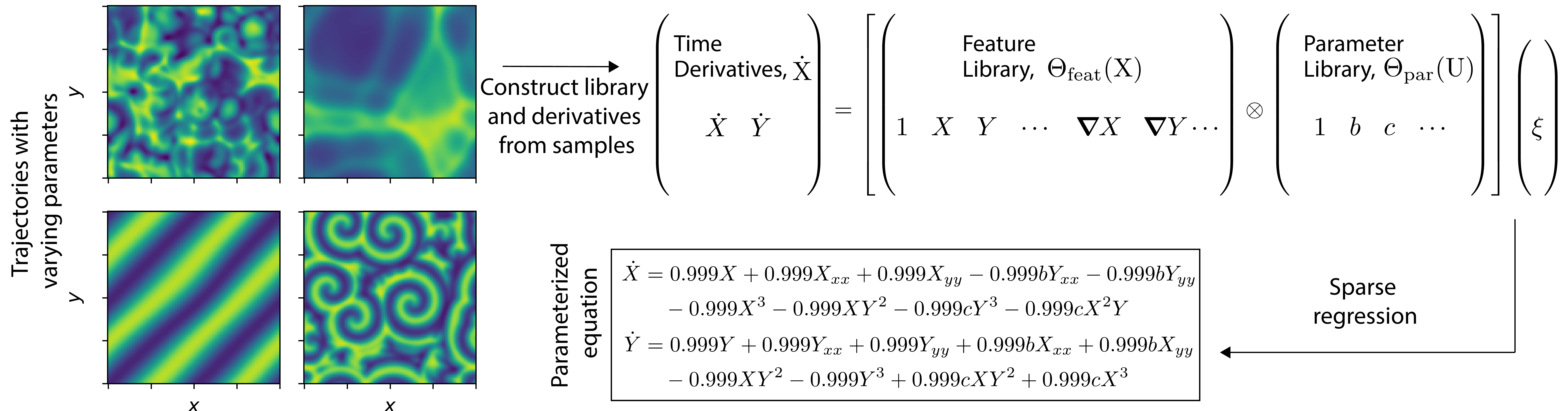}
\caption{Schematic of the SINDyCP approach. Data collected from sample trajectories collected under various driving parameters are processed to construct a matrix of time derivatives, a feature library $\Theta_{\mathrm{feat}}$ of possible governing terms, and a parameter library $\Theta_{\mathrm{par}}$ of parametric dependencies. Sparse regression is applied to the library coefficients $\xi$ to identify a  parameterized governing equation.  \label{fig1}}
\end{figure*}

Our approach is implemented in the open-source PySINDy repository \cite{2022_Kaptanoglu,github}, enabling other powerful methods to be used in conjunction (see Supplemental Material Sec.~S1A \cite{SM}). In particular, we develop and assess a weak formulation \cite{2020_Reinbold,2021_Reinbold,2021_Messenger,2022_Messenger} of SINDyCP, which shows excellent performance on noisy data. 
We demonstrate that the method can be easily and effectively employed to discover accurate parameterized models from the kind of data available in typical pattern-formation experiments and that these parameterized models enable extrapolation beyond the conditions under which they were developed.

\textit{Building the library.} Figure \ref{fig1} illustrates the SINDyCP approach applied to the spatiotemporal evolution of four trajectories of the complex Ginzburg-Landau equation
\begin{equation}
\dot{A} = A + (1+ib)\nabla^2 A - (1-ic)|A|^2A, \label{cgle}
\end{equation}
which is described by a complex dependent variable $A(\mathbf{x},t)$ in two spatial dimensions $\mathbf{x}=(x,y)$. Ginzburg-Landau exhibits a stunning variety of patterns, depending on the bifurcation parameters $b$ and $c$.  We generate four trajectories with parameter values ${(b,c)=(2.0,1.0), (2.0,0.75), (0.5,0.5)}$, and $(1.0,0.75)$, which exhibit differing dynamical phases, corresponding to amplitude turbulence, phase turbulence, stable waves, and frozen spiral glasses, respectively \cite{2002_Aranson}.  Our goal is to discover the partial differential equation for the real and imaginary components $A=X+iY$ parameterized by $b$ and $c$ from time series data.  

As with most SINDy algorithms, we first form a matrix of the input data $\mathrm{X}$, whose columns correspond to the dependent variables and whose rows correspond to the sample measurements of the dependent variables. In the case of Fig.~\ref{fig1}, for example, $\mathrm{X}$ consists of two columns corresponding to the real and imaginary parts of $A$ and $4N_xN_yN_t$ rows, where $N_x$, $N_y$, and $N_t$ are the number of sample points in the corresponding spatiotemporal dimensions; again, there are four trajectories. We then determine the temporal derivative $\dot{\mathrm{X}}$ for each sample point, either through numerical differentiation or through direct measurements.

In basic SINDy, we define a matrix of library terms $\Theta=\Theta(\mathrm{X})$ depending on the input data, which includes all possible terms that may be present in the differential equation that describes the temporal derivatives.  These terms may be built from polynomial combinations of the dependent variables and their spatial derivatives, for example, although more general libraries are possible. In the SINDYc approach, we augment the library dependence with an external control signal $\mathrm{U}$, i.e., $\Theta = \Theta(\mathrm{X},\mathrm{U})$. The library terms are typically determined by appending the control variables to the dependent variables and again forming polynomials and derivatives.  In the case in Fig.~\ref{fig1}, we can treat the parameters as external control signals, $\mathrm{U}=(b,c)$ and apply SINDYc, but the traditional implementation of this approach will fail for PDEs, as we show.

SINDYc aims to find a sparse linear combination of the library terms determined by the vector of coefficients $\xi$ which minimizes the fit error 
\begin{equation}
\xi^* = \mathrm{argmin}_\xi \left \lvert \dot{\mathrm{X}} - \Theta(\mathrm{X},\mathrm{U})\xi \right \rvert + \lambda \left \lvert\xi\right\rvert_0,
\end{equation}
where sparsity promoting regularizing term $\lambda \left \lvert\xi\right\rvert_0$ penalizes nonzero coefficients via the $L0$ norm. 
Crucially, all SINDy methods employ sparse regression (with appropriate regularization) to determine a sparse set of nonzero coefficients $\xi^*$. Such sparsity is expected in physically-relevant dynamics and produces parsimonious and interpretable models.  Here, we employ the sequentially thresholded least-squares algorithm \cite{2016_Brunton}, which iteratively eliminates library terms with coefficients that fall below a threshold hyperparameter.

One challenge that arises when applying the traditional SINDYc to control parameters in PDEs with existing implementations such as PySINDy is that the matrix of library terms $\Theta$ is traditionally formed by computing \textit{all}  polynomial combinations of spatial derivatives of the dependent and control variables.  However, since the control parameters are spatially constant,  the spatial derivatives will vanish identically, leading to a singular matrix $\Theta$. Such degeneracies lead to poor numerical results on data with control parameters when combining PDEFind and SINDYc without modification.  To overcome this challenge, we propose constructing a more general library through products of a feature library $\Theta_{\mathrm{feat}}(\mathrm{X})$ and a parameter library $\Theta_{\mathrm{par}}(\mathrm{U})$, as 
\begin{equation}
\Theta(\mathrm{X},\mathrm{U}) = \Theta_{\mathrm{feat}}(\mathrm{X})\otimes\Theta_{\mathrm{par}}(\mathrm{U}),
\end{equation}
where the product $\otimes$ here is defined to give the matrix consisting of all combinations of products of columns between the libraries, i.e., the $i$th row of $A\otimes B$ contains all the products of the form $A_{ij} B_{ik}$ where $j$ and $k$ span the columns of $A$ and $B$, respectively.  By distinguishing the feature and parameter library dependencies with this SINDyCP approach, we can construct much more targeted and well-conditioned libraries.

Using a feature library consisting of spatial derivatives up to third order and polynomials up to third order along with a linear parameter library, the SINDyCP approach easily discovers Eq.~\eqref{cgle} in Cartesian coordinates, as shown in Fig.~\ref{fig1}.  Additional demonstrations of SINDyCP for maps and ordinary differential equations (ODEs) are available in the Supplemental Material Sec.~S1B, and further details of the complex Ginzburg-Landau equation (CGLE) integration, along with an animation illustrating the temporal evolution of the sample trajectories, are available in the Supplemental Material Sec.~S2 \cite{SM}.

\textit{Amplitude dynamics beyond weakly nonlinear theory.} 
To illustrate the application of SINDyCP to pattern formation, we implement a numerical demonstration with the Belousov-Zhabotinksy chemical reaction system. We numerically integrate the Oregonator model \cite{1995_Mazzotti},
\begin{subequations}
\begin{align}
\dot{C}_X &= k_1 C_A C_H^2 C_Y-k_2C_HC_XC_Y+k_3C_AC_HC_X  \nonumber \\
&\quad-2k_4C_X^2  + D_X \nabla^2 C_X, \\
\dot{C}_Y &= -k_1C_AC_H^2C_Y-k_2C_HC_XC_Y + \nu k_5 C_B C_Z  \nonumber \\
&\quad+ D_Y \nabla^2 C_Y\\
\dot{C}_Z &= 2k_3C_AC_HC_X-k_5C_BC_Z + D_Z \nabla^2 C_Z,
\end{align}
\end{subequations}
which describes the evolution of oscillating chemical concentrations $C_X$, $C_Y$, and $C_Z$ for given supplied concentrations $C_A$, $C_B$, and $C_H$, and stoichiometric coefficient $\nu$, which depends on the experimental setup. We vary the concentration of $C_B$ and define a  control parameter $\mu \equiv (C_B -C_B^c)/C_B^c$, where $C_B^c$ is the critical value where the Hopf bifurcation occurs.   Section S3A of the Supplemental Material details the Oregonator model, along with a data-driven approach to detect and characterize the bifurcation point when the model is unknown \cite{SM}.  Here, we aim to develop a data-driven extension of  Eq.~\eqref{cgle} that incorporates nonlinear parameter dependence describing the dynamics far from the bifurcation. 
\begin{figure}[t]
\includegraphics[width=\columnwidth]{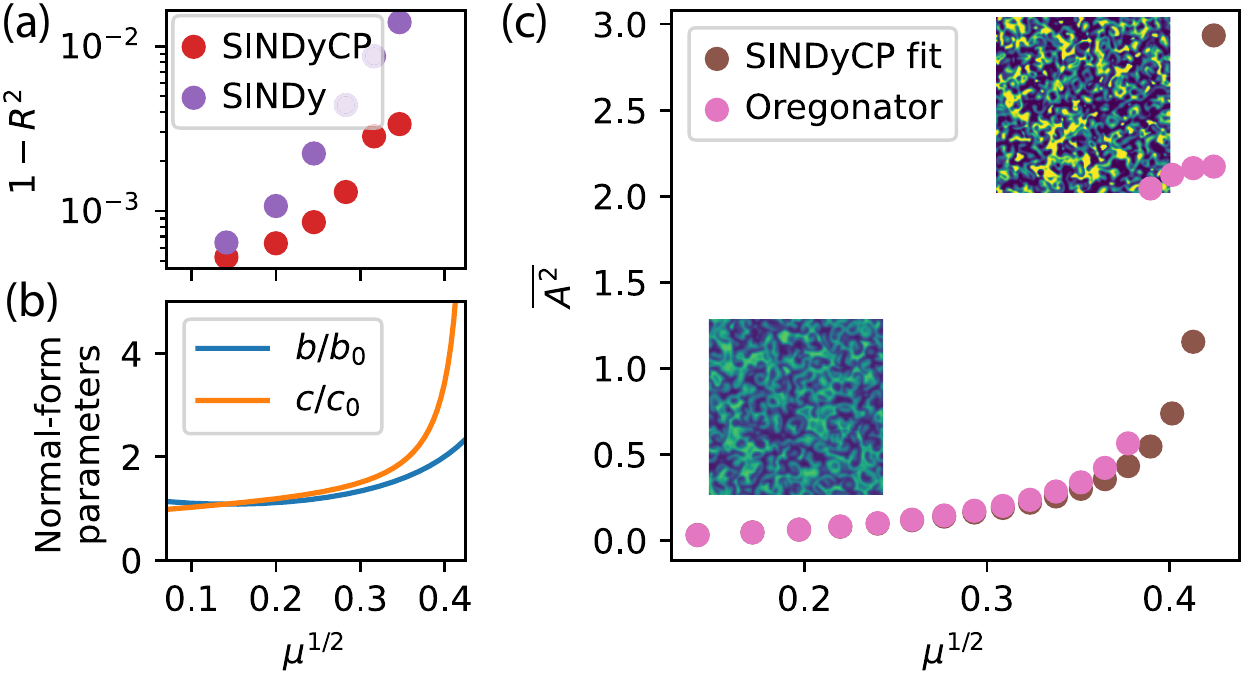}
\caption{Corrections to the weakly nonlinear theory of the Oregonator model. (a) $R^2$ score for the parameterized SINDyCP model and for an unparameterized SINDy fit on test trajectories collected at the parameter values used to train the model. (b) Corrected normal-form parameter values relative to the weakly nonlinear values $b_0$ and $c_0$ as a function of the bifurcation parameter $\mu^{1/2}$.  (c) Average limit cycle amplitude $\overline{A^2}$ vs $\mu^{1/2}$ for the Oregonator model and SINDyCP fit. The fit correctly exhibits a canard explosion far from the onset of the instability (insets show $C_X$ above and below the explosion, and an animation of the evolution is available in the Supplemental Material \cite{SM}).    \label{fig2}}
\end{figure}

We expect the dynamics near a Hopf point to be constrained to the two-dimensional center manifold, which describes the evolution of the complex amplitude dynamics governed by Eq.~\eqref{cgle}. 
When the governing equations are known, the weakly nonlinear theory develops a perturbative expansion near the Hopf point to express the complex amplitude $A$  in terms of the state space given by $\mathbf{x}\equiv(C_X-C_X^c, C_Y-C_Y^c, C_Z-C_Z^c)$. 
This theory follows from a near identity transformation of the governing equations up to cubic order,  as detailed in the Supplemental Material Sec.~S3B \cite{SM}. 

To demonstrate our approach, we develop a data-driven construction of the amplitude $A$ and its dynamical equations when the governing equations for the state space are unknown that is effective even far from the Hopf point. This approach follows from a SINDyCP fit on time series data for any two independent measurements of the state space (motivated by previous results \cite{1995_Mazzotti}, we choose to use $C_X$ and $C_Z$ here). Because we will employ the normal-form transformation below, we employ a cubic feature library with polynomial terms up to third-order and second-order spatial derivatives and a parameter library with polynomial terms up to second order for the control parameter $\mu^{1/2}$. Furthermore, we employ implicit  SINDy \cite{2016_Mangan} by including first-order temporal derivatives in the feature library.  Inverting the resulting implicit equations results in governing equations that are cubic in the state variables with coefficients that are rational functions of the control parameters, enabling the discovery of nonlinear corrections to parameter dependencies in the weakly nonlinear theory. Finally, by eliminating nonresonant coefficients using the normal-form transformation for cubic equations, we discover amplitude dynamics of the form in Eq.~\eqref{cgle}, but with normal-form coefficients $c(\mu)$ and $b(\mu)$ with rational dependence on the control parameter.  Additional details on the data-driven amplitude construction are available in the Supplemental Material Sec.~S3C \cite{SM}.

Figure \ref{fig2}(a) shows the $R^2$ score of the model on test trajectories corresponding to the parameter values that the model was trained on (a value of $R^2=1$ means that the fit perfectly predicts the temporal derivatives of the data). For reference, we also perform an unparameterized SINDy fit [purple dots in Fig.~\ref{fig2}(a)] on the training trajectory with the smallest $\mu$ value, which produces a model very close to the weakly nonlinear theory. The SINDyCP fit performs significantly better than the unparameterized fit, with $1-R^2$ nearly an order of magnitude smaller for the larger $\mu$ values.

The normal-form parameters $b(\mu)$ and $c(\mu)$ agree with the analytic values derived \cite{2000_Ipsen} from the original model as $\mu\to0$, but here we are able to discover them directly from data without any knowledge of the governing equations. Furthermore,  as shown in Fig. \ref{fig2}(b),  the variation of the parameters becomes extreme for $\mu^{1/2}>0.35$, which we were able to discover via the implicit version of SINDy. In fact, as shown in Fig. \ref{fig2}(c), the Oregonator model exhibits a canard explosion (in which the limit cycle amplitude expands abruptly due to highly nonlinear effects) \cite{1995_Mazzotti} around $\mu^{1/2}\approx 0.39$, where the weakly nonlinear theory breaks down. The SINDyCP model reflects this breakdown and enables the development new models that account for it.

\textit{Weak formulation.} The weak formulation utilizes integration against compactly supported ``test functions'' to defined the SINDy problem. The weak method shows excellent performance for noisy data, owing to its ability to minimize the need for computing numerical derivatives. Rather than forming samples (rows in Fig.~\ref{fig1}) from spatiotemporal points for each trajectory, the weak method constructs the system rows by projecting the data onto weak samples such as
\begin{equation}
w_{ik}^\nu \equiv \int_{\Omega_k} \phi (\mathbf{x}; t) \mathrm{X}_i^{(\nu)}(\mathbf{x}; t) ~{\mathrm d}^Dx{\mathrm d}t,
\end{equation}
where $\Omega_k$ is a compactly supported sample domain, $\phi$ is the test function, and $\mathrm{X}_i^{(\nu)}$ denotes the $\nu$th partial derivative the $i$th dependent variable. By moving derivatives off of the data and onto the test functions via integration by parts,
\begin{equation}
w_{ik}^\nu = (-1)^{|\nu|} \int_{\Omega_k} \phi^{(\nu)}(\mathbf{x}; t) \mathrm{X}_i (\mathbf{x}; t) ~{\mathrm d}^Dx{\mathrm d}t,
\end{equation}
the weak method significantly reduces the impact of measurement noise on the SINDy library and improves the fit results \cite{fn1}.

To maximize the performance of the weak method, we have optimized and fully vectorized numerical integration for the weak formulation in PySINDy, which can be easily combined with the SINDyCP library class.   Products of weak features do not generally form reasonable samples for a SINDy model, since multiplication and integration do not commute, so at first sight, it is not clear how to combine weak-form feature and parameter libraries with SINDyCP. However, when computing the weak samples corresponding to constant functions, such as those that form the parameter library, the integrals simply represent the spatiotemporal volume of the domain $\Omega_k$. Our implementation thus rescales the weak features for the temporal derivatives by the same volumetric factors. Details of our implementation are presented in Sec.~S4 of the Supplemental Material \cite{SM}. 

\textit{Performance.} Using $500$ randomly distributed sample domains (measuring $1/10$th the spatiotemporal domain size in each direction), the weak SINDyCP easily identifies the complex Ginzburg-Landau equation using the same data used for the traditional differential form shown in Fig.~\ref{fig1}. Furthermore, it can do so in just a few seconds of run-time on a modern processor in this case (over five times faster than the differential form).

\begin{figure}
\includegraphics[width=\columnwidth]{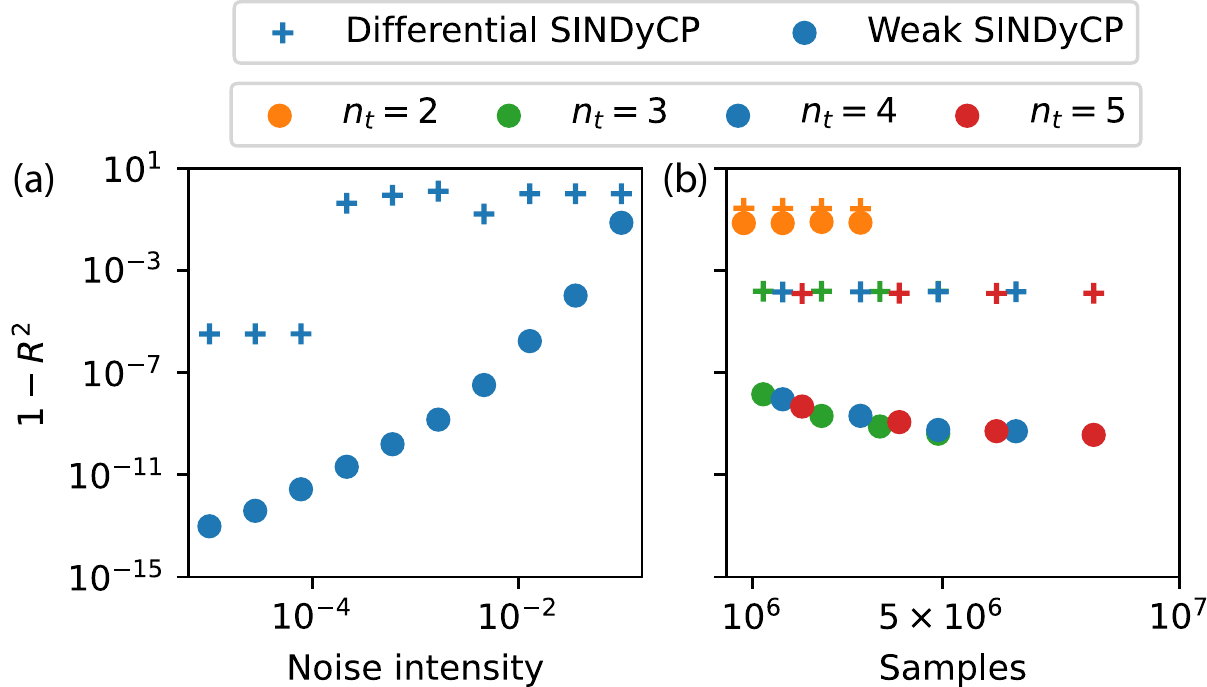}
\caption{Performance of SINDyCP for the fit of the complex Ginzburg-Landau equation with noisy data. (a) Model score vs noise intensity using the differential and weak forms of SINDyCP with $n_t=4$ trajectories. (b) Model score vs number of samples for varying number of randomly generated trajectories, varying trajectory length, and noise intensity $10^{-3}$. \label{fig3}}
\end{figure}

To assess the impact of noise, we inject random Gaussian noise of varying intensity \cite{fn2} into the four trajectories used as the training data for the complex Ginzburg-Landau equation. We then generate two new sample trajectories to use as testing data, with $b=2.0, 1.5$ and $c=1.5, 1.0$, respectively. Using the training data, we perform the SINDyCP fits using both the differential and weak formulation and evaluate the $R^2$ score on our test trajectories.  Figure \ref{fig3}(a) shows the results for the $R^2$ score on the test trajectories. While both methods provide good fits for low noise intensity, only the weak method exhibits a robust fit for parameterized equations for large noise intensities.

The SINDyCP fit also requires a sufficient amount of data to identify governing equations. Figure \ref{fig3}(b) shows the performance of SINDyCP on the testing data for fits performed with a varying number of trajectories $n_t=2,3,4,5$ and of varying length corresponding to a number of time samples $N_t=25,50,75,100$, with an injected noise intensity of $10^{-3}$. Unlike the trajectories in Fig.~\ref{fig1}, the parameters for trajectories were randomly generated, with $(b,c)$ distributed as Gaussian random variables with means $(1.5, 1.0)$ and standard deviations $(0.5,0.25)$. For too little data, the fit fails to identify the correct model, and the value of $1-R^2$ is $O(1)$. The models improve moderately with an increasing number of samples per trajectory (the product of $N_t$ with the number of spatial grid points). More importantly, a sufficiently large number of trajectories $n_t$ is required to achieve a good fit (at least $3$ in this case). The amount of data required will further increase when including a larger number of possible library terms and when identifying a larger number of parameters. These requirements should be carefully assessed in order to achieve successful SINDyCP fits for more general pattern-forming systems.

\begin{figure}
\includegraphics[width=\columnwidth]{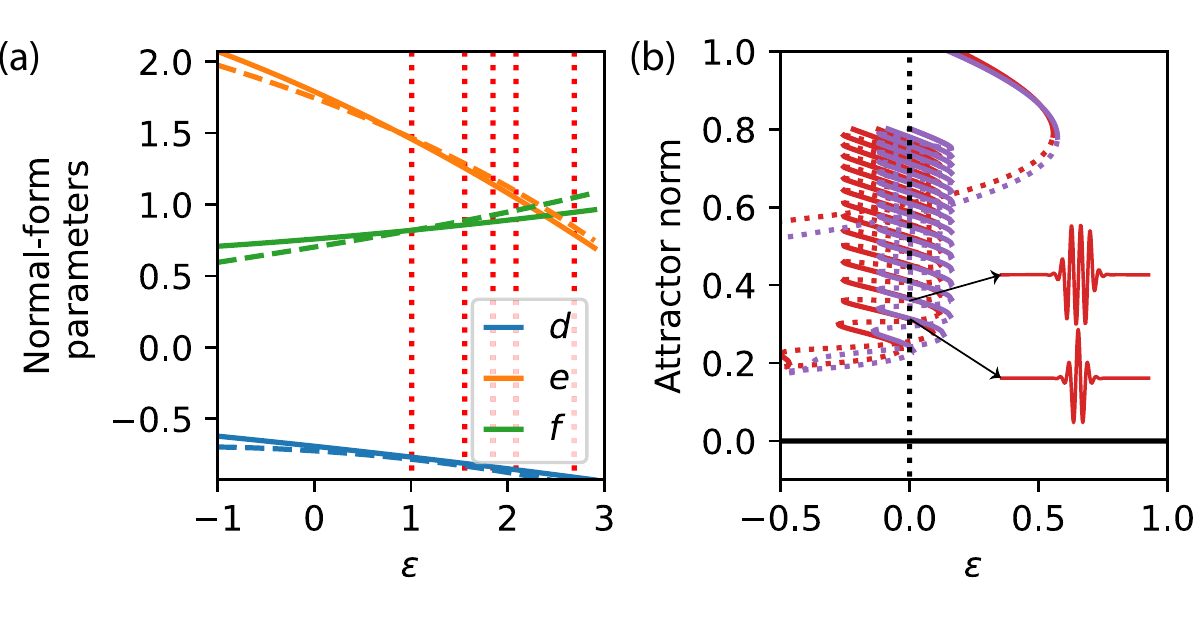}
\caption{Extrapolation of localized states in the cubic-quintic Swift-Hohenberg equation. (a) The randomly generated relationships between the normal-form parameters $(d,e,f)$ and the experimental parameter $\varepsilon$.  Red dotted lines show the values used to train the SINDyCP fit and dashed colored lines show the coefficients derived from the fit. (b) Snaking bifurcations of localized states for Eq.~\eqref{sh} (purple lines) and the SINDyCP fit (red lines). Insets show the localized states found from random initial conditions with $\varepsilon=0$ in the discovered model.  \label{fig4}}
\end{figure}

\textit{Parameter extrapolation.} As a final demonstration (Fig.~\ref{fig4}), we consider the one-dimensional cubic-quintic Swift-Hohenberg equation
\begin{equation}
\dot{u} = du - u_{xxxx} - 2u_{xx} -u +eu^3 -f u^5, \label{sh}
\end{equation}
with parameters $d$, $e$, and $f$ describing the linear,  cubic,  and regularizing quintic terms, respectively. This model pattern-formation equation has been used to study defect dynamics incorporating quintic corrections beyond the weakly nonlinear approximation and has revealed universal snaking bifurcations leading to the formation of localized states for $e>0$ and $d<0$ \cite{Burke_2007}.  

The parameters $d$, $e$, and $f$ are the minimal and natural set to describe the possible dynamics in the Swift-Hohenberg equation derived from normal-form theory. However, in typical pattern-formation applications, one does not have direct control over such parameters. Instead, experimentally accessible parameters will have a complicated and nonlinear relationship with the normal-form parameters, which requires detailed knowledge and tedious calculations to derive, e.g., an expansion and center manifold transformation around a bifurcation point. The SINDyCP approach enables an automated discovery of such relationships, which can be used to extrapolate system behavior beyond a set of measurements.

To illustrate this idea, we generate random quadratic relationships between an experimental parameter $\varepsilon$ and the normal-form parameters $(d,e,f)$, and we create three training trajectories using random values of the parameter $1<\varepsilon<3$ [Fig.~\ref{fig4}(a)].  For all of the training trajectories,  $\varepsilon$ is sufficiently large that no localized or periodic states exist, and all trajectories decay to the trivial $u=0$ solution. We perform the weak SINDyCP fit using these trajectories subject to injected white noise with intensity $\sigma=0.01$ \cite{fn2} and with a quadratic parameter library. To test the ability of SINDyCP to extrapolate beyond the parameter regime given in the input data, we simulate the identified model for the experimental parameter value $\varepsilon=0$. Remarkably, even with limited and noisy training data, the method identifies an accurate relationship between $\varepsilon$ and the normal-form parameters. Simulations of the identified model with random initial conditions converge to localized states for $\varepsilon=0$.  Numerical continuation of these localized states [Fig.~\ref{fig4}(b)] with the AUTO package \cite{2007_Doedel} reveals that the SINDyCP model exhibits snaking bifurcations that closely approximate those in the Swift-Hohenberg equation (see Sec.~S5 of the Supplemental Material \cite{SM} for fits and continuations with differing noise intensity and training data). Thus, despite the significant extrapolation of the parameter value beyond the input data, the model captures the complex bifurcation structure in the dynamics.

\textit{Discussion.} The SINDyCP approach represents a simple but powerful generalization of SINDy with control. By disambiguating the feature and parameter components of the SINDy libraries, the method enables the discovery of systems of partial differential equations parameterized by driving parameters. Such equations arise naturally in the context of pattern formation, where the normal forms of bifurcations lead to parameterized equations near the onset of instabilities. The approach can be easily applied with the data available in typical pattern-formation experiments and promises to enable extrapolation beyond the regime that can be theoretically described with weakly nonlinear theory.  For example, it may find application in the discovery of mechanisms leading to the formation of novel localized states beyond the snaking bifurcations of the Swift-Hohenberg equation \cite{2014_Chen,2021_Nicolaou}. While new phenomena may be easily conjectured to occur at unseen parameter values, we emphasize that such predictions must be validated experimentally to ensure correct extrapolation.

In practice, two significant challenges must be overcome to discover good parameterized models with SINDyCP. First, the method requires sufficiently informative trajectory data. Samples should be collected on appropriate temporal and spatial scales, sufficiently many parameter values should be measured, and trajectories with persistent dynamics provide more information than transient trajectories.  While the weak formulation significantly mitigates the problem, measurement noise impairs the fit and can corrupt results. Second, the method requires an appropriate state space with a good coordinate representation to discover sparse dynamics. Near a bifurcation, the normal-form theory helps provide information about the state space dimension and sparsity-promoting coordinate transformations. In the future, a more sophisticated data-driven phase-amplitude reconstruction \cite{2008_Kralemann} or autoencoder-assisted discovery of physical coordinates \cite{2019_Champion,2021_Smidt,2022_Chen,2022_Bakarji,2022_Cenedese} will further enable researchers to discover parsimonious equations governing complex systems directly from data gathered through experiments conducted under various driving parameters. 

\begin{acknowledgments}
This work benefited from insightful discussions with Alan Kaptanoglu. Z. G. N. is a WRF postdoctoral fellow.  We acknowledge support from the National Science Foundation AI Institute in Dynamic Systems (Grant No. 2112085). 
\end{acknowledgments}


\begin{thebibliography}{99}
\bibitem{book_bk} S. L. Brunton and J. N. Kutz, \textit{Data-driven science and engineering: Machine learning, dynamical systems, and control} (Cambridge University Press,  Cambridge, UK,  2022).
\bibitem{2020_Udrescu} S. M. Udrescu and M. Tegmark, AI Feynman: A physics-inspired method for symbolic regression, Sci. Adv. \textbf{6},  eaay2631 (2020).
\bibitem{2021_Karniadakis} G. E. Karniadakis, I. G. Kevrekidis, L. Lu, P. Perdikaris, S. Wang, and L. Yang, Physics-informed machine learning, Nat. Rev. Phys. \textbf{3}, 422 (2021).
\bibitem{2016_Brunton} S. L. Brunton, J. L. Proctor, and J. N. Kutz, Discovering governing equations from data by sparse identification of nonlinear dynamical systems, Proc. Natl. Acad. Sci. U.S.A. \textbf{113}, 3932 (2016).
\bibitem{1993_Cross} M. C. Cross and P. C. Hohenberg, Pattern formation outside of equilibrium, Rev. Mod. Phys. \textbf{65}, 851 (1993).
\bibitem{1998_Kuznetsov} Y. A. Kuznetsov, \textit{Elements of applied bifurcation theory} (Springer, New York, 1998).
\bibitem{2002_Aranson} I. S. Aranson and L. Kramer,The world of the complex Ginzburg-Landau equation, Rev. Mod. Phys. \textbf{74}, 99 (2002).
\bibitem{2017_Nicolaou} Z. G. Nicolaou, H. Riecke, and A. E. Motter, Chimera States in Continuous Media: Existence and Distinctness, Phys. Rev. Lett. \textbf{119}, 244101 (2017).
\bibitem{2022_Heinonen} V. Heinonen, A. J. Abraham, J. Słomka, K. J. Burns, P. J. S\'{a}enz, and J. Dunkel, Emergent universal statistics in nonequilibrium systems with dynamical scale selection,  arXiv:2205.01627 (2022).
\bibitem{2016_Brunton2} S. L. Brunton, J. L. Proctor, and J. N. Kutz, Sparse identification of nonlinear dynamics with control (SINDYc), IFAC-PapersOnLine \textbf{49}, 710-715 (2016).
\bibitem{2018_Kaiser} E. Kaiser, J. N. Kutz, and S. L. Brunton, Sparse identification of nonlinear dynamics for model predictive control in the low-data limit, Proc. Royal Soc. A \textbf{474}, 20180335 (2018).
\bibitem{2021_Fasel} U. Fasel, E. Kaiser, J. N. Kutz, B. W. Brunton, and S. L. Brunton, SINDy with control: A tutorial, in \textit{2021 60th IEEE Conference on Decision and Control (CDC)} (IEEE, Austin, 2021), pp. 16--21.
\bibitem{2017_Rudy} S. H. Rudy,  S. L. Brunton, J. L. Proctor,  and J. N. Kutz, Data-driven discovery of partial differential equations, Sci. Adv. \textbf{3}, e1602614 (2017).
\bibitem{2017_Schaeffer} H. Schaeffer, Learning partial differential equations via data discovery and sparse optimization, Proc. Royal Soc. A \textbf{473}, 20160446 (2017).
\bibitem{2019_Rudy} S. H. Rudy, A. Alla, S. Brunton,  and J. N. Kutz, Data-driven identification of parametric partial differential equations,  SIAM J. Appl. Dyn. Syst  \textbf{18},  643 (2019).
\bibitem{2017_Schaeffer_2} H. Schaeffer, G. Tran, and R. Ward, Learning dynamical systems and bifurcation via group sparsity, arXiv preprint arXiv:1709.01558 (2017).
\bibitem{2017_Kramer} B. Kramer, P. Grover, P. Boufounos, S. Nabi, and M. Benosman, Sparse sensing and DMD-based identification of flow regimes and bifurcations in complex flows, SIAM J.  Appl. Dyn. Syst. \textbf{16}, 1164 (2017).
\bibitem{2020_Schmelzer} M. Schmelzer, R. P. Dwight, and P.  Cinnella,  Discovery of algebraic Reynolds-stress models using sparse symbolic regression, Flow Turbul. Combust. \textbf{104},  579 (2020).
\bibitem{2020_Zanna} L. Zanna and T. Bolton, Data‐driven equation discovery of ocean mesoscale closures, Geophys. Res. Lett. \textbf{47}, e2020GL088376 (2020).
\bibitem{2021_Beetham} S. Beetham, R. O. Fox, and J. Capecelatro, Sparse identification of multiphase turbulence closures for coupled fluid–particle flows, J. Fluid Mech. \textbf{914}, A11 (2021).
\bibitem{2020_Wang} Z. Wang, B. Wu, K. Garikipati, and X. Huan, A perspective on regression and Bayesian approaches for system identification of pattern formation dynamics, Theor. Appl. Mech. Lett. \textbf{10}, 188-194 (2020).
\bibitem{2021_Romeo} N. Romeo, A. Hastewell, A. Mietke, and J. Dunkel, Learning developmental mode dynamics from single-cell trajectories, eLife \textbf{10}, e68679 (2021).
\bibitem{2021_Supekar}Supekar, Rohit, Boya Song, Alasdair Hastewell, Gary PT Choi, Alexander Mietke, and Jörn Dunkel. 
R. Supekar, B. Song, A. Hastewell, G.P.T. Choi,  A. Mietke, and J. Dunkel, Learning hydrodynamic equations for active matter from particle simulations and experiments, Proc. Natl. Acad. Sci.  U. S. A. \textbf{120},  e2206994120 (2023).
\bibitem{2022_Kaptanoglu}A. A. Kaptanoglu \textit{et al.},PySINDy: A comprehensive Python package for robust sparse system identification, J. of Open Source Softw. \textbf{7}, 3994 (2022).
\bibitem{github}The PySINDy repository is available at \href{https://github.com/dynamicslab/pysindy}{https://github.com/dynamicslab/pysindy}.
\bibitem{2020_Reinbold} P. A. K. Reinbold, D. R. Gurevich, and R. O. Grigoriev, Using noisy or incomplete data to discover models of spatiotemporal dynamics, Phys. Rev. E \textbf{101}, 010203(R) (2020).
\bibitem{2021_Reinbold} P. A. K. Reinbold, L. M. Kageorge, M. F. Schatz, and R. O. Grigoriev, Robust learning from noisy, incomplete, high-dimensional experimental data via physically constrained symbolic regression, Nat. Comm. \textbf{12}, 3219 (2021).
\bibitem{2021_Messenger} D. A. Messenger and D. M. Bortz, Weak SINDy for partial differential equations, J. of Comput. Phys. \textbf{443}, 110525 (2021).
\bibitem{2022_Messenger} D. A. Messenger and D. M. Bortz, Learning mean-field equations from particle data using WSINDy, Physica D \textbf{439}, 133406 (2022).
\bibitem{SM} See Supplemental Material at \url{http://link.aps.org/supplemental/10.1103/PhysRevResearch.5.L042017} for an animation of the trajectories in Figs.~\ref{fig1} and \ref{fig2} and details about numerical integration, additional demonstrations, the Oregonator model, the weak form implementation, and the Swift-Hohenberg equation.
\bibitem{1995_Mazzotti}M. Mazzotti, M. Morbidelli, and G. Serravalle,  Bifurcation analysis of the Oregonator model in the 3-D space bromate/malonic acid/stoichiometric coefficient, J. Phys. Chem. \textbf{99},  4501 (1995).
\bibitem{2016_Mangan}N. M. Mangan, S. L. Brunton, J. L.  Proctor, and J.  N. Kutz, Inferring biological networks by sparse identification of nonlinear dynamics,IEEE Trans. Mol. Biol. Multi-Scale Commun.  \textbf{2}, 52 (2016).
\bibitem{2000_Ipsen} M. Ipsen, F.  Hynne, and P.  G. S{\o}rensen, Amplitude equations for reaction–diffusion systems with a Hopf bifurcation and slow real modes, Physica D \textbf{136}, 66 (2000).
\bibitem{fn1}It is not possible to remove all numerical derivatives in the weak formulation, but the maximum order of derivatives can generally be reduced to at most half the original order for the library.
\bibitem{fn2}Noise intensity here refers to the pointwise standard deviation on the spatiotemporal grid employed in the simulations. True white noise has a Dirac delta variance, and intensity should thus scale with grid spacing and time step to 1/2 power.
\bibitem{Burke_2007} J. Burke and E. Knobloch, Homoclinic snaking: structure and stability,Chaos \textbf{17}, 037102 (2007).
\bibitem{2007_Doedel} E. J. Doedel, A. R. Champneys, F. Dercole, T. F. Fairgrieve, Y. A. Kuznetsov, B. Oldeman, R. C. Paffenroth, B. Sandstede, X. J. Wang, and C. H. Zhang, AUTO-07P: Continuation and bifurcation software for ordinary differential equations, \href{http://indy.cs.concordia.ca/auto/}{http://indy.cs.concordia.ca/auto/}.
\bibitem{2014_Chen} B. G. Chen, N. Upadhyaya, and V. Vitelli, Nonlinear conduction via solitons in a topological mechanical insulator, Proc. Natl. Acad. Sci. U.S.A. \textbf{111}, 13004 (2014).
\bibitem{2021_Nicolaou} Z. G. Nicolaou, D. J. Case, E. B. Wee, M. M. Driscoll, and A. E. Motter, Heterogeneity-stabilized homogeneous states in driven media, Nat. Comm. \textbf{12}, 4486 (2021).
\bibitem{2008_Kralemann} B. Kralemann, L. Cimponeriu, M. Rosenblum, A. Pikovsky, and R. Mrowka, Phase dynamics of coupled oscillators reconstructed from data, Phys. Rev. E \textbf{77}, 066205 (2008).
\bibitem{2019_Champion} K. Champion, B. Lusch, J. N. Kutz, and S. L. Brunton, Data-driven discovery of coordinates and governing equations, Proc. Natl. Acad. Sci. U.S.A. \textbf{116}, 22445 (2019).
\bibitem{2021_Smidt}T. E. Smidt, M. Geiger, and B. K. Miller, Finding symmetry breaking order parameters with Euclidean neural networks, Phys. Rev. Research \textbf{3}, L012002 (2021).
\bibitem{2022_Chen} B. Chen, K. Huang, S. Raghupathi, I. Chandratreya, Q. Du, and H. Lipson, Automated discovery of fundamental variables hidden in experimental data, Nat. Comput. Sci. \textbf{2}, 433 (2022).
\bibitem{2022_Bakarji} J. Bakarji, K. Champion, J. N. Kutz, and S. L. Brunton, Discovering governing equations from partial measurements with deep delay autoencoders, Proc. Royal Soc. A \textbf{479}, 20230422 (2023).
\bibitem{2022_Cenedese}M. Cenedese,  J. Ax\r{a}s,  B. B\"{a}uerlein,  K. Avila, and G. Haller, Data-driven modeling and prediction of non-linearizable dynamics via spectral submanifolds, Nat. Comm. \textbf{13}, 872 (2022).






\end{thebibliography}
\end{document}


\renewcommand{\tiny}{\fontsize{6}{7}\selectfont}
\renewcommand{\scriptsize}{\fontsize{8}{9.5}\selectfont}
\renewcommand{\footnotesize}{\fontsize{9}{11}\selectfont}
\renewcommand{\small}{\fontsize{10}{12}\selectfont}
\renewcommand{\normalsize}{\fontsize{10.95}{13.6}\selectfont}
\renewcommand{\large}{\fontsize{12}{14}\selectfont}
\renewcommand{\Large}{\fontsize{14}{18}\selectfont}
\renewcommand{\LARGE}{\fontsize{17.28}{22}\selectfont}
\renewcommand{\huge}{\fontsize{20.74}{25}\selectfont}
\renewcommand{\Huge}{\fontsize{24.88}{30}\selectfont}
\setlength{\parindent}{15pt}
\renewcommand{\thefigure}{{\small S\arabic{figure}}}
\renewcommand{\thesection}{{\normalsize S\arabic{section}}}
\renewcommand{\theequation}{{\normalsize S\arabic{equation}}}
\normalsize
\onecolumngrid

{\center \Large \textbf{Supplementary Material for ``Data-Driven Discovery and Extrapolation of Parameterized Pattern-Forming Dynamics''} \\[1em]
\normalsize Zachary G. Nicolaou, Guanyu Huo, Yihui Chen, Steven L. Brunton, and J. Nathan Kutz \\[2em] }

\section{Software and source code}
\subsection{Installation}
We recommend installing the PySINDy package from source by cloning the GitHub repository with git to obtain the most up-to-date version.  We highly recommend doing so with a new Anaconda environment in order to avoid unexpected dependency conflicts. Anaconda can be installed following the instructions here: \url{https://www.anaconda.com/download#downloads}. The following bash commands will download the PySINDy repository, create an anaconda environment, and install the package:
\begin{lstlisting}[language=bash] 
cd $HOME
git clone https://github.com/dynamicslab/pysindy.git
cd $HOME/pysindy
conda create -y -n pysindy_env pip Jupyter matplotlib
conda activate pysindy_env
pip install -e .
\end{lstlisting}

Once installed, the easiest way to develop code is in a Jupyter notebook, and the repository provides several example notebooks in the examples subdirectory of the package.  We recommend starting with the first example notebook, which provides an overview of all the features implemented in the package.  Running the commands
\begin{lstlisting}[language=bash]
cd $HOME/pysindy/examples
jupyter notebook 1_feature_overview.ipynb
\end{lstlisting}
will launch a browser with the notebook, and running all the cells in sequence will demonstrate the package.  Finally, the developers currently actively maintain the package, and issues with the code can be addressed on the GitHub page: \url{https://github.com/dynamicslab/pysindy/issues}.

\subsection{Demonstrations}
All the results in this manuscript can be reproduced by sequentially running all cells in the 17th example notebook, which can be launched with
\begin{lstlisting}[language=bash]
cd $HOME/pysindy/examples/17_parameterized_pattern_formation
jupyter notebook parameterized_pattern_formation.ipynb
\end{lstlisting}
\begin{figure}[h!]
\includegraphics[width=\columnwidth]{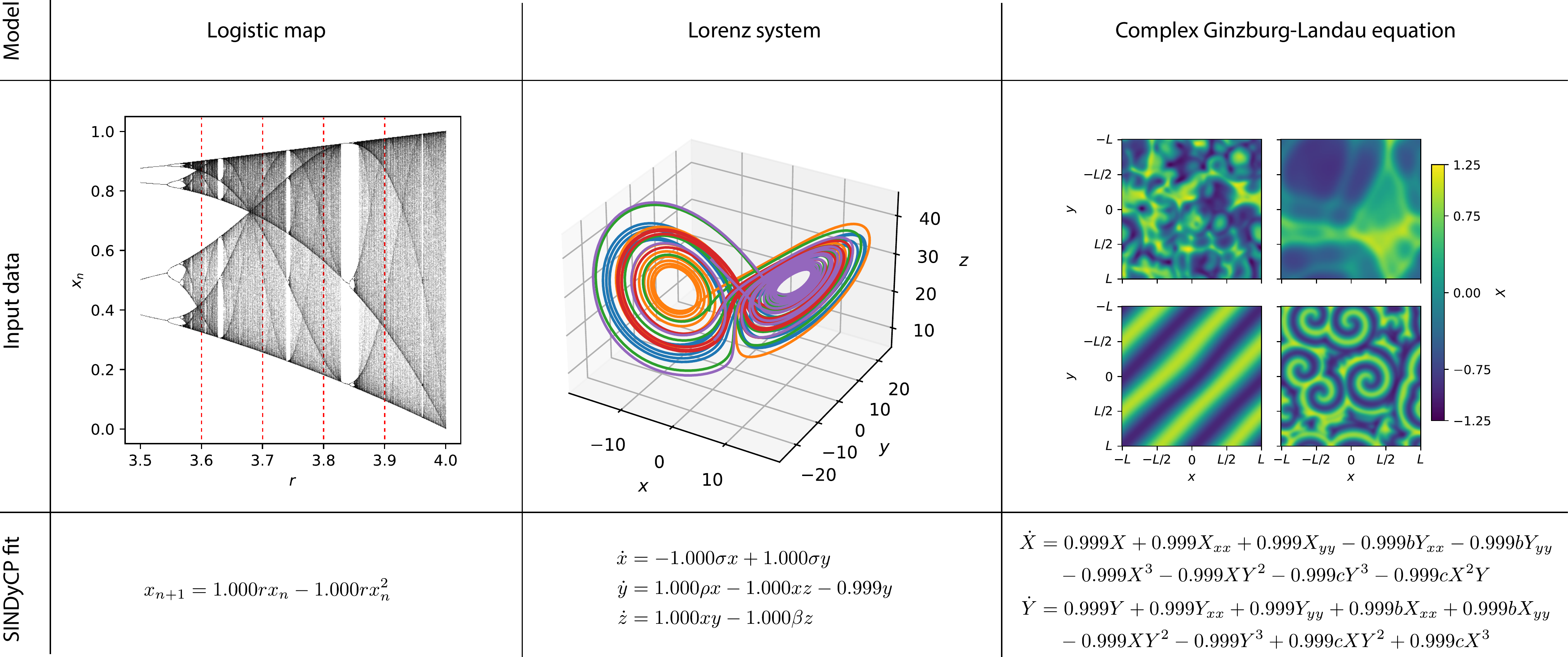}
\caption{Demonstrations of the SINDyCP approach for three models (top row) of nonlinear dynamics. Several trajectories produced from different parameter values (middle row) are supplied as input, and the SINDyCP fit (bottom row) correctly identifies the governing equations in each case. \label{figs2}}
\end{figure}
Note that since PDE data is high-dimensional, some of the cells require up to 100GB of RAM to run successfully, so we recommend running the notebook on a cluster server if available or downsampling the input.

In addition to the results in the main text, the notebook illustrates simpler pedagogical demonstrations of SINDyCP in discrete maps, ODEs, and PDEs shown in Fig.~\ref{figs2}.  The left panels illustrate the logistic map,
\begin{equation}
x_{n+1}=rx_n(1-x_n),
\end{equation}
which is a discrete-time system with a single dependent variable $x_n$ and a single parameter $r$. This equation is the model for a universal period-doubling route to chaos as the parameter $r$ increases past $3.56995$. We perform the SINDyCP fit using four sample trajectories of $1000$ iterations, corresponding to parameter values $r=3.6,3.7,3.8,3.9$ (red dotted lines in Fig.~\ref{figs2}). We employ a library consisting of polynomials up to third order in the dependent variable $x_n$ and linear functions of the control parameter $r$, and the SINDyCP approach correctly identifies the parameterized equation. The middle panels illustrate the Lorenz system,
\begin{align}
\dot{x} = \sigma (y - x),
\dot{y} = x(\rho - z) - y, 
\dot{z} = x y - \beta z,
\end{align}
which consists of three ordinary differential equations in three dependent variables $x$, $y$, and $z$ and three parameters $\sigma$, $\rho$, and $\beta$. This equation exhibits the iconic butterfly-shaped Lorenz attractor for certain parameter values. We perform the SINDyCP fit using five sample trajectories that have converged to their attractors, corresponding to the randomly selected parameter values $\sigma=10.0, 9.8, 9.9, 10.3, 9.5$, $\rho=27.6, 28.2, 28.3, 27.6, 28.1$, and $\beta =3.1, 2.4, 2.4, 2.3, 2.4$, respectively. We use feature and parameter libraries consisting of polynomials up to fourth order in the dependent variables $(x,y,z)$ and linear functions in the parameters $(\sigma, \rho, \beta)$, and the SINDyCP approach again correctly identifies the parameterized equation. Finally, the right panels illustrate the CGLE described in the main text. 

\section{Complex Ginzburg-Landau Equation}
\begin{figure}[hb]
\includegraphics[width=0.5\columnwidth]{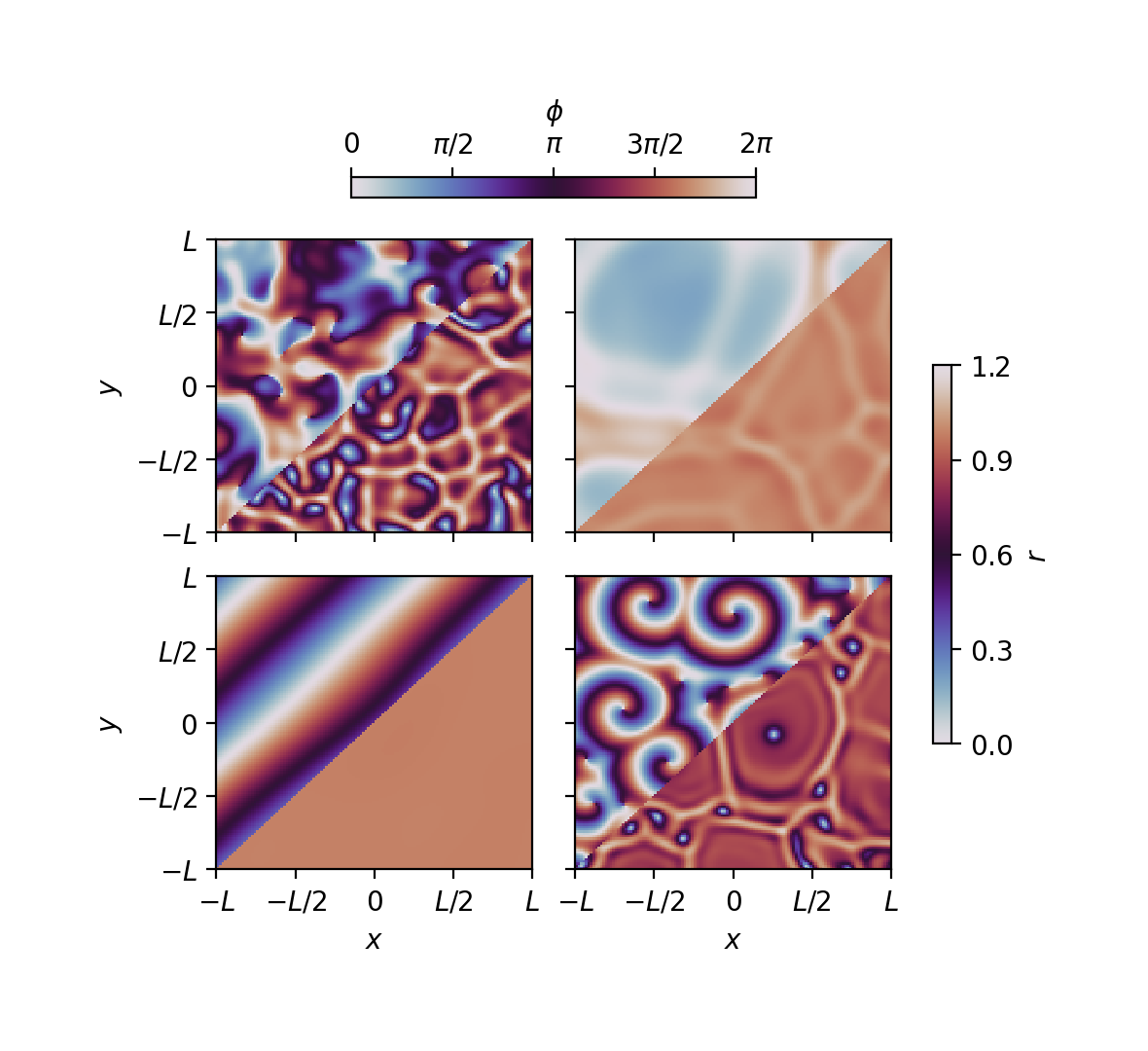}
\caption{Snapshot of the animation showing the complex amplitude $A=re^{i\phi}$ of the trajectories.  In each panel, the phase $\phi$ is shown in the upper left and the amplitude $r$ is shown in the lower right. \label{figs1}}
\end{figure}

To integrate the complex Ginzburg-Landau equation, we use a pseudospectral integration method. We take a periodic domain of size of size $L=32\pi$ in each direction and discretize using $N_x=N_y=128$ grid points in each spatial direction. Derivatives are calculated using fast Fourier transforms, and the discretized system is integrated with a 4(5) Runge-Kutta-Fehlberg method (which is also used for the other equations, with relative and absolute error tolerances of $10^{-6}$). To produce states in the dynamical phases of interest, we take random initial conditions $A_0 = \sum_{nm} \alpha_{nm}  e^{i\mathbf{k}_{nm}\cdot \mathbf{x}} + \epsilon e^{i\mathbf{k}_{2\,2}\cdot \mathbf{x}}$, where $\alpha_{nm}$ are complex random Gaussian amplitudes with mean zero and variance $\sigma^2/(1+n^2+m^2)$, $\mathbf{k}_{nm} = 2\pi(n\hat{\mathbf{x}}+m\hat{\mathbf{y}})/L$, the sum ranges over $-2\leq n,m \leq 2$, and $\epsilon$ is the scale of an initial plane wave perturbation with wavevector $k_{2\,2}$.  The mode amplitudes are determined by $\sigma=0.1,0.1,0.1,1.0$ and $\epsilon=0.01,0.01,1.0,0.01$ for the four trajectories used in the main text. The system is allowed to approach an attractor for the first $90$ time units, then the trajectory is formed by the next $10$ time units, in steps of $0.01$. We also provide an animation showing the phase and amplitude for longer runs of $100$ time units (Fig.~\ref{figs1}). 

\section{Oregonator model and weakly nonlinear theory}
\subsection{Model details and bifurcation identification}
\label{hopfsec}
We mainly follow the analyses of the Oregonator model in Refs.~[31, 33], with realistic parameter values shown in Table \ref{tabs1}.  The fixed point $(C_X,C_Y,C_Z)=(C_X^0, C_Y^0, C_Z^0)$ undergoes a Hopf bifurcation as the control concentration $C_B$ decreases below the critical value $C_B^c=0.787$, leading to oscillatory chemical dynamics.  While this is known for the Oregonator model, here we demonstrate that this bifurcation can be predicted and quantified if the governing equations were unknown using the SINDyCP approach.
\begin{table}[b!]
\begin{tabular}{c | c | c | c | c | c | c | c | c | c | c | c}
$k_1$ & $k_2$ & $k_3$ & $k_4$ & $k_5$ & $D_X$ & $D_Y$ & $D_Z$ & $C_H$ & $C_A$ & $C_B^c$ & $\nu$ \\
\hline 
2 & $10^6$ & $10$ & $2 \times 10^3$ & $1$ &$10^{-5}$ & $1.6 \times 10^{-5}$ & $0.6 \times 10^{-5}$ & $0.5$ & $1$ & $0.787$ & $1$
\end{tabular}
\caption{Parameter values for the Oregonator model, in cgs units (suppressed for brevity).\label{tabs1}}
\end{table}

We initially fix a spatial domain of length $L=1/\sqrt{0.05} \text{ cm}\approx4.47 \text{ cm}$ and employ the pseudospectral integration with an integration time of $T=400\text{ s}$, and we produce trajectories for $C_B=(0.826, 0.834, 0.842, 0.850, 0.858,0.866)$. We record the concentrations for time steps of $dt=0.594804 \text{ s}$, which corresponds to $1/10$th the period of the critical frequency at $\mu=1$.  Anticipating the two-dimensional state space required to describe a Hopf bifurcation, we consider fits for the variables $(C_X, C_Z)$.  Each of these trajectories decays to a fixed point $(C_X^0(\mu), C_Z^0(\mu))$, and we estimate the fixed point for each $C_B$ by averaging the trajectories for the final $100$ time steps.  

\begin{figure}[b]
\includegraphics[width=0.9\linewidth]{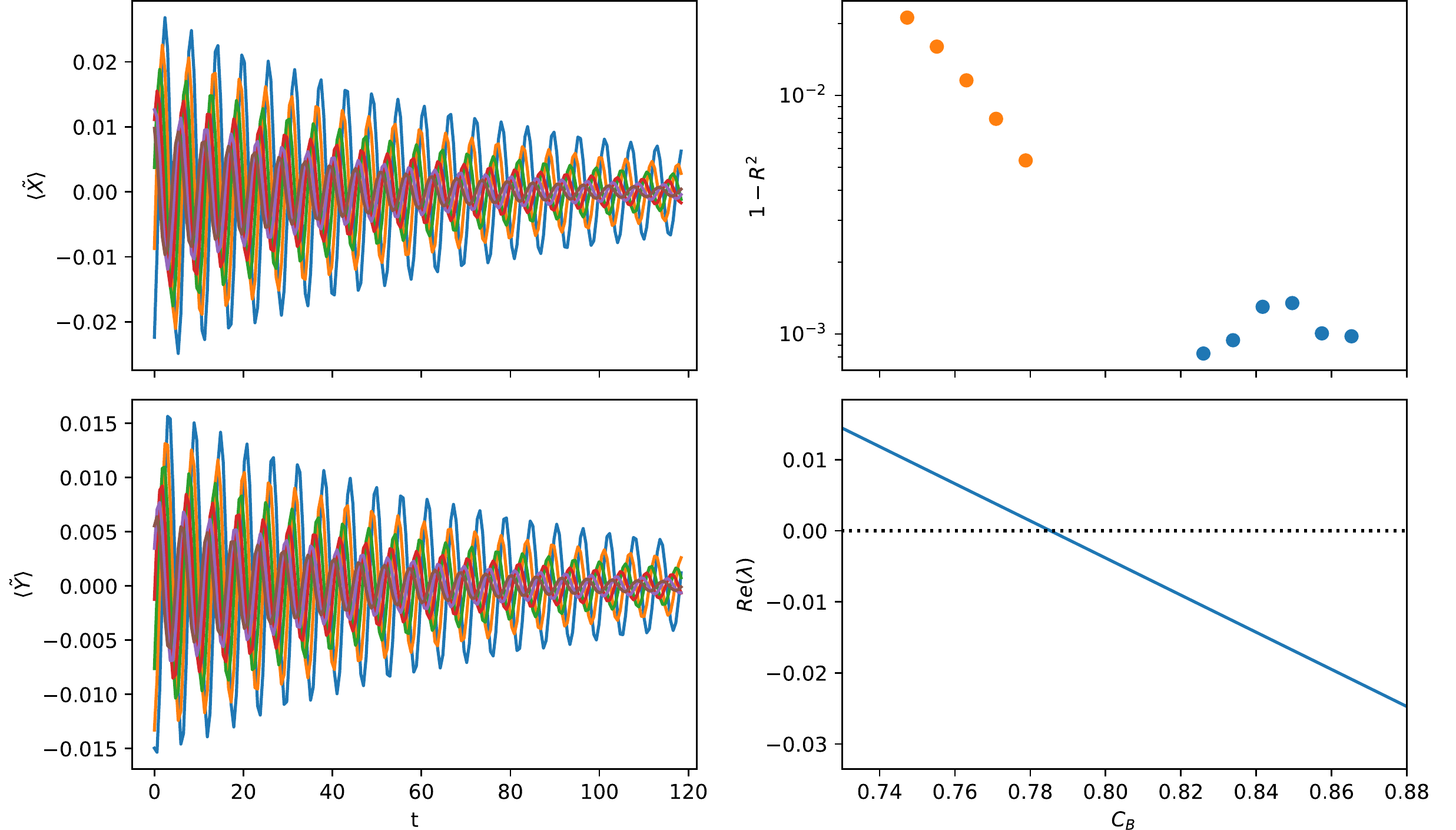}
\caption{Estimation of the Hopf point from data.  The left two panels show the spatially-averaged scaled coordinates as a function of time; the top-right panel shows the $R^2$ score for test data corresponding to the training parameters (blue dots) and for extrapolation to beyond the Hopf point (orange dots); and the bottom-right panel shows the real part of the linear spectrum as a function of $C_B$. \label{figs3}}
\end{figure}

We then construct scaled coordinates 
\begin{equation}
\left(\tilde{X},\tilde{Y}\right) \equiv \left(\frac{C_X-C_X^0}{C_X^0}, \frac{C_Z-C_Z^0}{C_Z^0}\right) \label{scaled}
\end{equation}
to fit with SINDyCP. We use the first 200 time steps as training data and the next 200 time steps as test data; Fig.~\ref{figs3}(a) shows the spatial average $\left(\langle \tilde{X} \rangle,  \langle\tilde{Y} \rangle \right) \equiv \int \left(\tilde{X},\tilde{Y}\right)  \mathrm{d}x \mathrm{d}y /L^2$ of the training data sets. The dynamics are dominated by exponential decay,  so we use a linear feature library.   Furthermore, since we assume no knowledge of the location of the bifurcation, we assume a linear parameter library for the control parameter $C_B$ as well.   The SINDyCP fit produces a model
\begin{align}
\dot{\tilde{X}} &=  (0.374+0.455C_B)\tilde{X}+(-1.832-0.292C_B)\tilde{Y},\\
\dot{\tilde{Y}} &=  (-0.035+0.977C_B)\tilde{X}+(0.028-0.968C_B)\tilde{Y}.
\end{align}
The stability of the constant solution to this model is determined by the eigenvalues $\lambda$ of the Jacobian matrix $J = \begin{pmatrix} 0.374+0.455C_B && -1.832-0.292C_B \\   -0.035+0.977C_B && 0.028-0.968C_B \end{pmatrix}$.   Figure \ref{figs3} shows the real part of the eigenvalues as a function of $C_B$. As expected, the spectrum is stable for the training data. The spectrum of the fit becomes unstable as $C_B$ decreases past $C_B=0.784$, closely predicting the true instability of $C_B^c=0.787$ as well as the critical frequency.  
To assess the ability to extrapolate the dynamics, we also produce additional testing trajectories for $C_B=(0.748, 0.756, 0.763, 0.771, 0.779)$ and evaluated the $R^2$ score of the model (orange dots in Fig.~\ref{figs3}).  The model performs poorly, as expected given the strictly linear dynamics. More sophisticated fitting does not resolve this issue (see Section \ref{ampsec} below), since the transient decay in the training data is too fast to produce a good estimate for the dynamics beyond the Hopf bifurcation. While the Hopf point can be easily predicted in this case, good quantification of the dynamics requires training past the Hopf point.

\subsection{Weakly nonlinear theory}
\label{weaklynonlinearsec}
Define the bifurcation parameter $\mu=(C_B^c-C_B)/C_B^c$, as in the main text. For small $\mu$, the weakly nonlinear theory follows from a perturbative expansion of the model. Take $\mathbf{x} \equiv (C_X,C_Y,C_Z)-(C_X^0, C_Y^0, C_Z^0)$ and express Eqs.~(4)-(6) as $\dot{\mathbf{x}}=\mathbf{F}(\mathbf{x})$. Define the multilinear operators of partial derivatives $\mathbf{F}_{\mathbf{x}^n}(\mathbf{e}_{i_1},\cdots,   \mathbf{e}_{i_n}) = {\partial^n \mathbf{F}}/{\partial x_{i_1}\cdots\partial x_{i_n}}\rvert_{\mathbf{x}=0}$ with $\mathbf{e}_{i}$ the $i$th component unit vector.  Noting that $\mathbf{x}=O(\mu^{1/2})$,  the Taylor expansion for the system is 
\begin{align}
\dot{\mathbf{x}} &=  D\cdot \nabla^2\mathbf{x} + \mathbf{F}_{\mathbf{x}^1}(\mathbf{x})\rvert_{\mu=0} + \left(\partial \mathbf{F}/\partial \mu\rvert_{\mu=0}\right) \mu  + \frac{1}{2} \mathbf{F}_{\mathbf{x}^2}(\mathbf{x},\mathbf{x})\rvert_{\mu=0} \nonumber\\
&\quad+  \left(\left(\partial \mathbf{F}/\partial \mu\right)_{\mathbf{x}^1}(\mathbf{x})\rvert_{\mu=0}\right)\mu + \frac{1}{6} \mathbf{F}_{\mathbf{x}^3}(\mathbf{x},\mathbf{x},\mathbf{x})\rvert_{\mu=0}+\cdots, \label{taylor}
\end{align}
where $D$ is a diagonal matrix with elements $D_X$, $D_Y$ and $D_Z$.  For such cubic dynamics, the normal-form theory develops a near-identity transformation $\mathbf{x}=\mathbf{y} + \mathbf{h}(\mathbf{y},\mu)$ perturbatively. 
After a short transient, the dynamics relax onto the center manifold, where 
\begin{equation}
\mathbf{y} \equiv A e^{i\omega_0 t} \mathbf{u}  + \bar{A} e^{-i\omega_0 t} \bar{\mathbf{u}}. \label{ycoord}
\end{equation}
Here $\mathbf{u}$ is one of the critical eigenvectors of the Jacobian matrix $\mathbf{F}_{\mathbf{x}^1}$ with eigenvalue $\lambda = i\omega_0$ (with zero real part for $\mu=0$) and overbars represent complex conjugates, and we also define the corresponding left eigenvector at $\mathbf{u}^\perp$.   
The near-identity transformation function $\mathbf{h}(\mathbf{y},\mu)$ is selected to eliminate the non-resonant terms in the evolution equation of $A$, which can be accomplished under general conditions. This results in an amplitude equation $\dot{A} = \mu \sigma A + g|A|^2A + d\nabla^2 A$,
where 
\begin{align}
\sigma &= \mathbf{u}^\perp \cdot \left(\partial \mathbf{F}/\partial \mu \right)_{\mathbf{x}^1}(\mathbf{u})- \mathbf{u}^\perp \cdot \mathbf{F}_{\mathbf{x}^2}\left[\mathbf{u},\left(\mathbf{F}_{\mathbf{x}^1}\right)^{-1}\left(\partial \mathbf{F}/\partial \mu \right)\right], \\
g&=\frac{1}{2}\mathbf{u}^\perp \cdot \mathbf{F}_{\mathbf{x}^3}\left(\mathbf{u},\mathbf{u},\bar{\mathbf{u}}\right)-\mathbf{u}^\perp \cdot \mathbf{F}_{\mathbf{x}^2}\left\{\mathbf{u},\left[\mathbf{F}_{\mathbf{x}^1}\right]^{-1}\left[\mathbf{F}_{\mathbf{x}^2}\left(\mathbf{u},\bar{\mathbf{u}}\right)\right]\right\} \nonumber \\
&\quad -\frac{1}{2}\mathbf{u}^\perp \cdot \mathbf{F}_{\mathbf{x}^2}\left\{\bar{\mathbf{u}},\left[\mathbf{F}_{\mathbf{x}^1}-\left(\lambda-\bar{\lambda}\right)I\right]^{-1}\left[\mathbf{F}_{\mathbf{x}^2}\left(\mathbf{u},\mathbf{u}\right)\right]\right\}, \label{normalg}\\
d&=\mathbf{u}^\perp \cdot D \cdot \mathbf{u}. \label{normald}
\end{align}
By rescaling the amplitude by a factor of $\mu^{1/2}$, time by a factor of $1/\mu$, and space by a factor of $1/\mu^{1/2}$ and employing additional rescalings to unitize the real components and eliminate the mean rotation, we can arrive at the CGLE in Eq.~(1) of the main text, where $b\equiv\mathrm{Im}(d)/\mathrm{Re}(d)=b_0=0.173$ and $c\equiv-\mathrm{Im}(g)/\mathrm{Re}(g)=c_0=2.379$.  As expected, these parameter values correspond to the amplitude turbulence regime of the CGLE.

\subsection{Data-driven amplitude construction and normal-form parameters}
\label{ampsec}
\begin{figure}[t]
\includegraphics[width=0.9\linewidth]{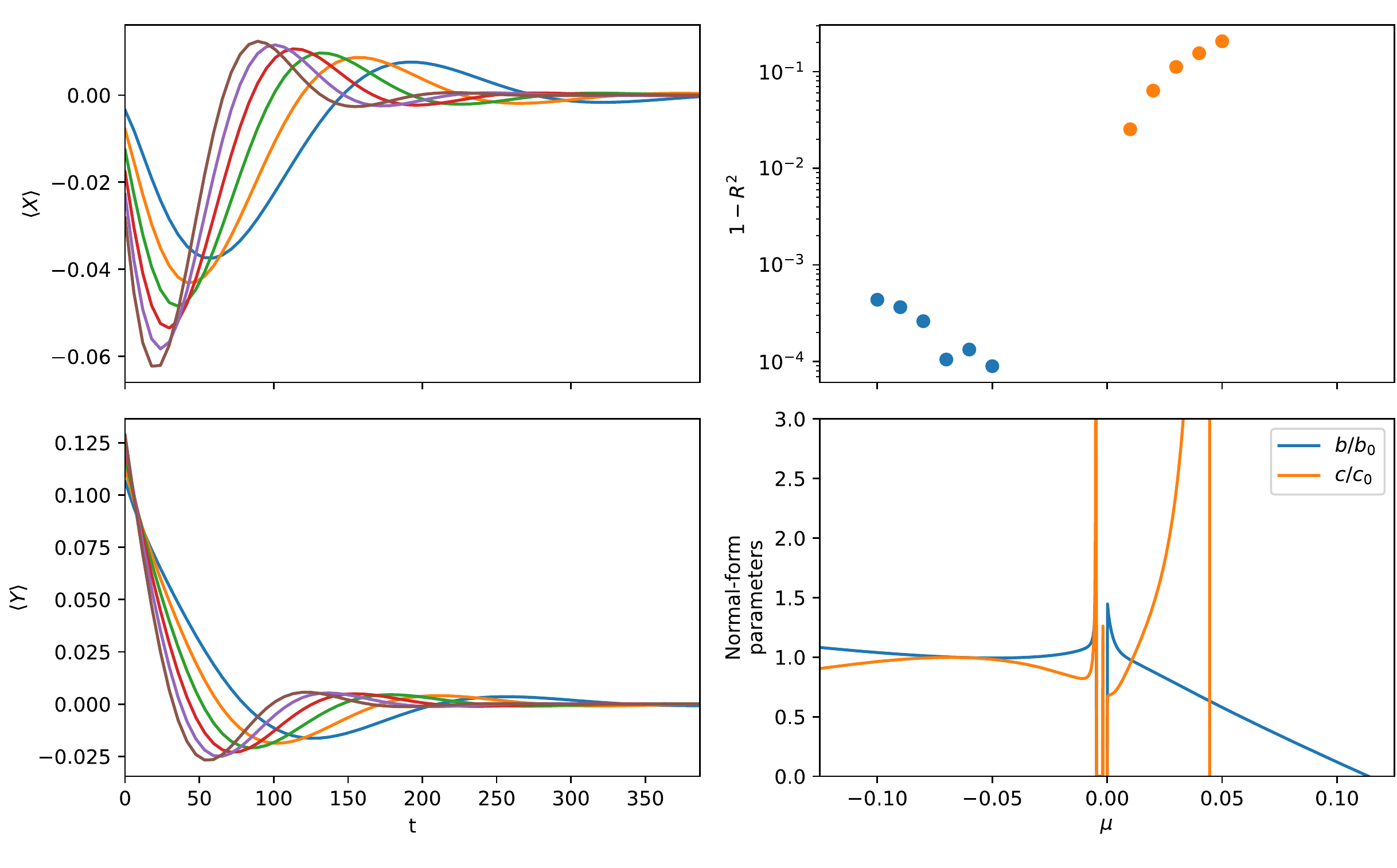}
\caption{SINDyCP fits for the $\mu<0$ data. The left two panels show the spatially-averaged amplitudes as a function of time; the top-right panel shows the $R^2$ score for test data in the corresponding to the training parameters (blue dots) and for extrapolation to $\mu>0$ test data (orange dots); and the bottom-right panel shows the predicted normal-form parameters as a function of $\mu$. \label{figs4}}
\end{figure}

Motivated by Eq.~\eqref{ycoord} in the weakly nonlinear theory, we can generate a data-driven amplitude $A=X+iY$ by averaging the scaled coordinates in Eq.~\eqref{scaled} over a period of the fast oscillation in a rotating reference frame
\begin{equation}
A \equiv \int_t^{t+2\pi/\omega_0} \left(\tilde{X}+i\tilde{Y}\right) e^{-i\omega_0 t} dt. \label{dataamp}
\end{equation}
This simple amplitude construction is superior to naive strobing at the critical period and works well for our purposes, but we note that more sophisticated approaches employing, e.g., analytic signal theory [40],  may be helpful in more complex scenarios. We rescale each trajectory by the time- and space-averaged amplitude $\int |A|  \mathrm{d}t  \mathrm{d}x  \mathrm{d}y $ in order to weigh each trajectory equally in the SINDyCP fit. Figure \ref{figs4} shows the spatially-averaged amplitude for the trajectories with $\mu<0$ corresponding to those in Section \ref{hopfsec}.  As desired, the time averaging removes the fast oscillations from the data and produces a smooth envelop evolution. Given the expected scaling of the Hopf bifurcation, we include a parameter library with square root and linear features and a feature library with polynomial features up to cubic terms and spatial derivatives up to second order. The equations discovered by SINDyCP with a threshold of $0.1$ are
\begin{align}
\dot{X}&=
(0.191-1.975\sqrt{\lvert\mu\rvert}+8.690\mu)X+(0.163-1.542\sqrt{\lvert\mu\rvert}+25.604\mu)Y\nonumber\\
&+(-1.628+11.824\sqrt{\lvert\mu\rvert}+18.764\mu)XX+(0.834-4.941\sqrt{\lvert\mu\rvert}+3.051\mu)XY\nonumber\\&+(-1.356+12.331\sqrt{\lvert\mu\rvert}+24.188\mu)YY\nonumber\\
&+(-24.868+126.916\sqrt{\lvert\mu\rvert}+281.099\mu)XXX+(33.179-80.714\sqrt{\lvert\mu\rvert}-118.407\mu)XXY\nonumber\\&+(-16.636+65.413\sqrt{\lvert\mu\rvert}+141.665\mu)XYY+(28.160-44.357\sqrt{\lvert\mu\rvert}-74.515\mu)YYY\nonumber\\
&+(1.214-0.344\sqrt{\lvert\mu\rvert}-0.677\mu)X_{xx}+(1.231-0.484\sqrt{\lvert\mu\rvert}-0.957\mu)X_{yy}\nonumber\\
&+(0.350-1.390\sqrt{\lvert\mu\rvert}-3.459\mu)Y_{xx}+(0.349-1.361\sqrt{\lvert\mu\rvert}-3.382\mu)Y_{yy}\\
\dot{Y}&=
(-0.246+2.401\sqrt{\lvert\mu\rvert}-22.813\mu)X+(14.423\mu)Y\nonumber\\
&+(-0.981+8.051\sqrt{\lvert\mu\rvert}+14.930\mu)XX+(-0.862+4.977\sqrt{\lvert\mu\rvert}+8.072\mu)XY\nonumber\\&+(0.456-1.047\sqrt{\lvert\mu\rvert}+12.951\mu)YY\nonumber\\
&+(-27.408+40.025\sqrt{\lvert\mu\rvert}+55.135\mu)XXX+(-14.839+29.917\sqrt{\lvert\mu\rvert}+15.043\mu)XXY\nonumber\\&+(-17.396-33.201\sqrt{\lvert\mu\rvert}-72.045\mu)XYY+(-15.098+37.491\sqrt{\lvert\mu\rvert}+24.892\mu)YYY\nonumber\\
&+(-0.161-0.411\sqrt{\lvert\mu\rvert}-1.294\mu)X_{xx}+(-0.172-0.320\sqrt{\lvert\mu\rvert}-1.089\mu)X_{yy}\nonumber\\
&+(1.177+0.233\mu)Y_{xx}+(1.178+0.272\mu)Y_{yy}
\end{align}
where the subscripts denote the spatial derivatives. The model performs very well for the parameter regime that it was trained under, with a $R^2$ score greater than $0.999$, but the dynamics for $\mu<0$ are too transient to give enough information to build a model that can extrapolate well. Thus, the model does not extrapolate very well, producing poor $R^2$ scores for test trajectories with $\mu>0$. 

The weakly nonlinear CGLE results in Section~\ref{weaklynonlinearsec} rely on the Taylor series expansion Eq.~\eqref{taylor} for $\mu$ near zero, but the normal-form transformation eliminating non-resonant terms applies more generally to asymptotic series up to cubic order in the state variables with arbitrary parameter dependence (see, e.g. Lemma 3.7 in Ref. [6]).  The values of the normal-form parameters under this general transformation remain $b(\mu)=\mathrm{Im}(d)/\mathrm{Re}(d)$ and $c(\mu)=-\mathrm{Im}(g)/\mathrm{Re}(g)$ where $g$ and $d$ are defined as in Eq.~\eqref{normalg}-\eqref{normald} as before, but now with $\mu$-dependent values of $\mathbf{F}_{\mathbf{x}^n}$. Consistently, the normal form parameters for the fits with $\mu<0$ closely approximate the analytic results $b(0)\approx b_0$ and $c(0)\approx c_0$ in the training regime (Fig.~\ref{figs4}), but significant variations emerge for $\mu>0$. As noted previously, the transient trajectories for $\mu<0$ do not contain enough information to build a model that extrapolates well.

For the $\mu>0$ trajectories from the main text, we use a scaled domain of size $L=1/\sqrt{\mu}$ and integration time $T=500/\mu$ according to the expectations from the Hopf bifurcation to obtain informative training trajectories. The amplitudes $A$ are then calculated and interpolated onto a common temporal grid with $500$ time points. The first fifth of the trajectory is discarded as the initial conditions relax to the chaotic attractor, the next two-fifths are used as training data, and the final two-fifths are used as test data.  

For the smallest value of $\mu=0.01$, we perform an unparameterized SINDy fit with a threshold of $0.1$ resulting in
\begin{align}
\dot{X}&=
1.402X+4.184Y\nonumber\\
&-0.694XXX+1.881XXY-0.664XYY+1.887YYY\nonumber\\
&+2.374X_{xx}+2.375X_{yy}+0.419Y_{xx}+0.419Y_{yy},\\
\dot{Y}&=
-4.219X+1.390Y\nonumber\\
&-1.877XXX-0.791XXY-1.893XYY-0.763YYY\nonumber\\
&-0.475X_{xx}-0.474X_{yy}+2.365Y_{xx}+2.365Y_{yy}.
\end{align}
The normal-form transformation gives parameter values $b=0.189$ and $c=2.586$, which approximate the weakly nonlinear theory values of $b_0=0.173$ and $c_0=2.379$ fairly well (this fit can be improved with additional trajectory data). Thus, the data-driven amplitude construction works well enough to reproduce the results of weakly nonlinear theory even when the governing equations are unknown.

\begin{figure}[hb]
\includegraphics[width=0.75\linewidth]{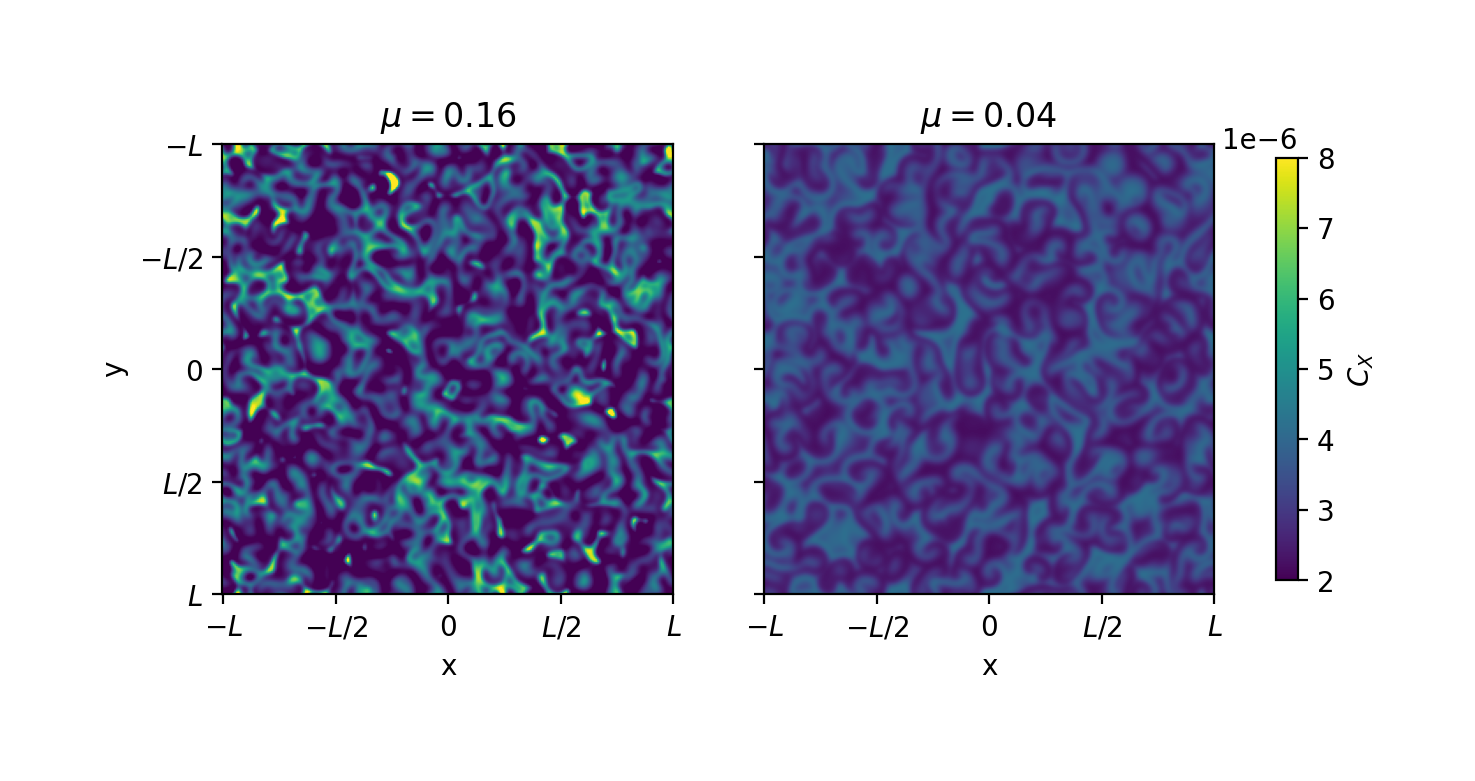}
\caption{Snapshot of the animation showing the concentration $C_X$ above (left) and below (right) the canard explosion.  \label{figs9}}
\end{figure}

For the parameterized SINDyCP fit in the main text, we employ the implicit SINDy to fit, which includes additional features columns for $\dot{X}$ and $\dot{Y}$,  leading to in highly nonlinear rational coefficients when solving for explicit governing equations.  The resulting SINDyCP fit with a threshold of $0.1$ is shown in Eqs.~\eqref{oregonatorfit1}-\eqref{oregonatorfit2} below. This model extrapolates much better than the weakly nonlinear theory.   Further sparsity is achieved by applying the normal-form transformation, leading to the parameter-dependent values of $b(\mu)$ and $c(\mu)$ [main text Fig.~2(b)].  We also provide an animation contrasting the pattern formation above and below the canard explosion (Fig.~\ref{figs9}).
\begin{align}
\dot{X}&=
(-0.699\sqrt{\lvert\mu\rvert}+1.836\mu)\dot{X}+(0.176\sqrt{\lvert\mu\rvert}-1.223\mu) \nonumber\\
&+(1.421+1.622\sqrt{\lvert\mu\rvert}-6.617\mu)X+(4.602-2.399\sqrt{\lvert\mu\rvert}+9.326\mu)Y\nonumber\\
&+(-0.436\sqrt{\lvert\mu\rvert}+3.911\mu)XX+(0.139-1.600\sqrt{\lvert\mu\rvert}+4.660\mu)XY\nonumber\\&+(-0.329\sqrt{\lvert\mu\rvert}+1.719\mu)YY\nonumber\\
&+(-0.838-0.372\sqrt{\lvert\mu\rvert}+6.856\mu)XXX+(1.564+5.179\sqrt{\lvert\mu\rvert}-13.666\mu)XXY\nonumber\\&+(-0.880+7.880\mu)XYY+(1.542+5.631\sqrt{\lvert\mu\rvert}-15.494\mu)YYY\nonumber\\
&+(2.127+5.220\sqrt{\lvert\mu\rvert}-15.257\mu)X_{xx}+(2.136+5.144\sqrt{\lvert\mu\rvert}-15.104\mu)X_{yy}\nonumber\\
&+(0.600-1.285\sqrt{\lvert\mu\rvert}+1.970\mu)Y_{xx}+(0.590-1.192\sqrt{\lvert\mu\rvert}+1.768\mu)Y_{yy} \label{oregonatorfit1}\\
\dot{Y}&=
(-0.760\sqrt{\lvert\mu\rvert}+2.089\mu)\dot{Y}+(0.350\mu)\nonumber\\
&+(-4.586+2.172\sqrt{\lvert\mu\rvert}-10.841\mu)X+(1.174+4.301\sqrt{\lvert\mu\rvert}-13.855\mu)Y\nonumber\\
&+(-0.177\sqrt{\lvert\mu\rvert}+1.359\mu)XX+(-0.151+1.976\sqrt{\lvert\mu\rvert}-7.154\mu)XY\nonumber\\&+(0.447\sqrt{\lvert\mu\rvert}-3.970\mu)YY\nonumber\\
&+(-1.494-6.473\sqrt{\lvert\mu\rvert}+18.922\mu)XXX+(-0.643-2.559\sqrt{\lvert\mu\rvert}+7.459\mu)XXY\nonumber\\&+(-1.612-4.896\sqrt{\lvert\mu\rvert}+12.900\mu)XYY+(-0.701-1.652\sqrt{\lvert\mu\rvert}+5.452\mu)YYY\nonumber\\
&+(-0.470-2.163\mu)X_{xx}+(-0.468-2.183\mu)X_{yy}\nonumber\\
&+(2.052+6.277\sqrt{\lvert\mu\rvert}-18.883\mu)Y_{xx}+(2.046+6.344\sqrt{\lvert\mu\rvert}-19.044\mu)Y_{yy}
\label{oregonatorfit2}
\end{align}

\section{Weak formulation implementation}
We refer the reader to Refs. [26-29] for the theory of the weak formulation of SINDy. Here, we only briefly describe our efficient numerical integration method for the weak formulation used in PySINDy. We suppose that the spatial grid is one-dimensional, for the moment, and the values of the coordinates on the grid points are $x_i$. The weak form requires us to calculate the integral of interpolated data $f(x)$ weighted by the $d$th derivatives of test function $\phi(x)$,
\begin{equation}
    I^{(d)} \equiv \int_{x_0}^{x_N}f(x)\phi^{(d)}(x)dx.
\end{equation} 
We choose to use test functions $\phi(x)=(x^2-1)^p$ in our implementation (after centering and rescaling the spatial coordinates so that $x_0=-1$ and $x_N=1$; we ignore the rescaling coefficients here), and thus their $d$th derivatives are
\begin{align}
    \phi^{(d)}(x) &= \frac{\partial}{\partial{x^d}}(x^2-1)^p = \sum_{k=0}^{p}\begin{pmatrix} p \\ k \end{pmatrix}(-1)^k \frac{(2(p-k))!}{(2(p-k) -d)!} x^{2(p-k)-d}.
\end{align}

We are provided with some feature values $u_i$ at the grid points, and we consider the value of a library function $f$ applied to that feature, $f_i \equiv f(u_i)$. We linearly interpolate the function as $f(x) = f_i + \frac{x-x_i}{x_{i+1}-x_i}(f_{i+1}-f_i)$ where $x_i \leq x \leq x_{i+1}$.  
Expanding the interpolation, and integrating the $x\phi^{(d)}(x)$ terms by parts,
\begin{align}
    I^{(d)} &= \sum_{i=0}^{N-1}\int_{x_i}^{x_{i+1}}\left[f_i + \frac{x-x_i}{x_{i+1}-x_i}(f_{i+1}-f_i) \right] \phi^{(d)}(x)dx \nonumber \\
    &= \sum_{i=0}^{N-1} \frac{f_i x_{i+1}-f_{i+1}x_i}{x_{i+1}-x_i}\left[\Phi^{(d)}(x_{i+1})-\Phi^{(d)}(x_i) \right]\nonumber \\ 
    &+ \frac{f_{i+1}-f_i}{x_{i+1}-x_{i}}\left[\Phi^{(d-1)}(x_{i+1})-\Phi^{(d-1)}(x_i) \right], \label{sum}
    \end{align}
where $\Phi^{(d)}(x)$ are the antiderivatives of $\phi^{(d)}$ [i.e. $\Phi^{(d)}(x) = \phi^{(d-1)}(x)$ for $d>0$].

By relabelling the dummy summation variables, we can recast Eq.~\eqref{sum} as a dot product between the input data $f_j$ and a weight $w_j$
\begin{align}
    I^{(d)} &= \sum_{j=0}^{N-1} w_j \cdot f_j, \label{dot}
\end{align}
with
\begin{align}
    w_j &\equiv \frac{x_{j+1}\left[\Phi^{(d)}(x_{j+1})-\Phi^{(d)}(x_j) \right]}{x_{j+1}-x_j} - \frac{x_{j-1}\left[\Phi^{(d)}(x_{j})-\Phi^{(d)}(x_{j-1}) \right]}{x_{j}-x_{j-1}} \nonumber \\
    &+ \frac{\Phi^{(d-1)}(x_j)-\Phi^{(d-1)}(x_{j-1})}{x_j - x_{j-1}} -
    \frac{\Phi^{(d-1)}(x_{j+1})-\Phi^{(d-1)}(x_{j})}{x_{j+1} - x_{j}}, \label{iweights}
\end{align}
where $0<j<N-1$. At the left and right sides of the domain (for $j=0$ and $j=N-1$), we must adjust the weights to correct for boundary effects,
\begin{align}
     w_0 &\equiv \frac{x_{1}\left[\Phi^{(d)}(x_{1})-\Phi^{(d)}(x_0) \right]}{x_{1}-x_0} -
    \frac{\Phi^{(d-1)}(x_{1})-\Phi^{(d-1)}(x_{0})}{x_{1} - x_{0}}, \label{lweights} \\
    w_{N-1} &\equiv  - \frac{x_{N-2}\left[\Phi^{(d)}(x_{N-1})-\Phi^{(d)}(x_{N-2}) \right]}{x_{N-1} - x_{N-2}} + \frac{\Phi^{(d-1)}(x_{N-1})-\Phi^{(d-1)}(x_{N-2})}{x_{N-1} - x_{N-2}}. \label{rweights}
\end{align}

Expressing the integrals along each dimension as dot products [Eq.~\eqref{dot}] enables efficient vectorization with BLAS operations, and the integration weights [Eqs.~\eqref{iweights}-\eqref{rweights}] are evaluated (in a vectorized fashion) just one time when the library is first initialized. We further vectorize the code by forming tensor products over all integration dimensions to calculate multidimensional integrals using a single tensor dot product.

\section{Swift Hohenberg equation and localized states}
As noted in the main text, we randomly generated a quadratic relationship between the ``experimental'' parameter $\varepsilon$ and the normal-form parameters $d(\varepsilon)$, $e(\varepsilon)$ and $f(\varepsilon)$.  Specifically, we take  $d(\varepsilon)=\sum_{n=0}^2 d_n \varepsilon^n$, $e(\varepsilon)=\sum_{n=0}^2 e_n \varepsilon^n$, and $f(\varepsilon)=\sum_{n=0}^2 f_n \varepsilon^n$ with $d_n$, $e_n$ and $f_n$ sampled uniformly in the intervals $[-10^{-n},10^{-n}]$, $[-5\times10^{-n},5\times10^{-n}]$, and $[0,10^{-n}]$, respectively.  The resulting equation is
\begin{align}
\dot{u}&=-2 u_{xx}-u_{xxxx}+(-1.691-0.073\varepsilon-0.003\varepsilon^2)u\nonumber\\
&+(1.791-0.306\varepsilon-0.025\varepsilon^2)u^3
-(0.758+0.056\varepsilon+0.005\varepsilon^2)u^5, \label{originalsh}
\end{align}
which we treat as ground truth. We selected these specific intervals for random sampling simply to localize the snaking bifurcations around $\varepsilon=0$, as shown in Fig. ~4(b) of the main text. 

As before,  we employ a pseudospectral method to integrate the Swift-Hohenberg equation, with $N_x=256$ discretization points, a domain of size $L=64 \pi$, an integration time of $5$ time units, and random initial condition is given by the real part of $u_0=\sum_{n=-20}^{20} \alpha_n e^{ik_nx}$ with $k_n=2\pi n/L$ and $\alpha_n$ complex random Gaussian amplitudes with mean zero and variance $1.0/(1+\sqrt{|n|})^2$. We randomly selected five values from a uniform distribution on $\varepsilon\in(1,3)$ and generated high-$\varepsilon$ trajectories well above the snaking bifurcations, and we similarly selected another five values on $\varepsilon\in(-2,1)$ to generate low-$\varepsilon$ trajectories around the snaking bifurcations.  Additionally, we added noise sampled from a Gaussian distribution with variance $\sigma^2$ and mean zero to produce the training data.
\begin{figure}[b]
\includegraphics[width=0.9\linewidth]{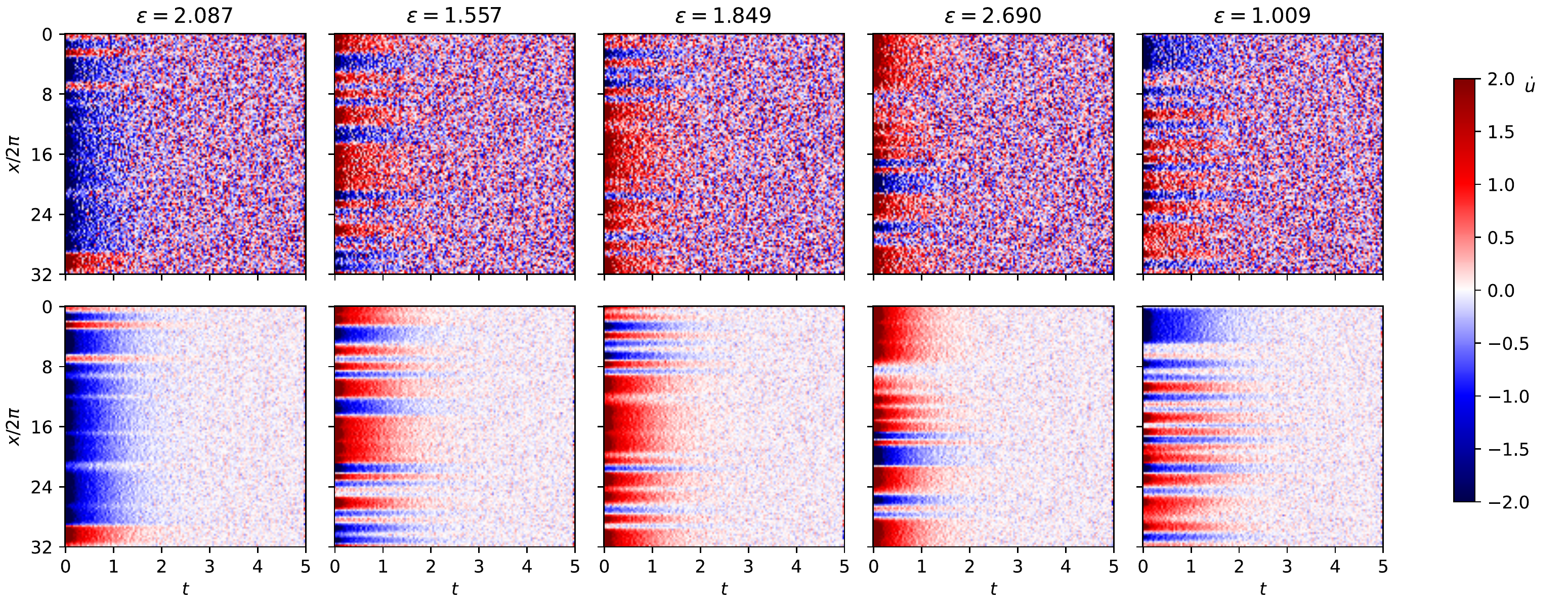}
\caption{Temporal derivatives for the high-$\varepsilon$ noisy data in the Swift-Hohenberg equation with $\sigma=0.05$ (top row) and $\sigma=0.01$ (bottom row). \label{figs5}}
\end{figure}

To maintain the reflection symmetry that ensures that homoclinic orbits can be continued under one-parameter continuation, we use a feature library consisting of odd-order polynomials up to order 5 and even spatial derivatives up to fourth order. This choice is simply a matter of convenience; when the reflection symmetry is broken slightly, localized states persist but begin to travel, which complicates the numerical continuation. On the high-$\varepsilon$ training trajectories shown in Fig.~\ref{figs5}, the weak SINDyCP produces a fit
\begin{align}
\dot{u}&=(-1.940-0.072\varepsilon+0.022\varepsilon^2)u_{xx}
+(-0.975-0.025\varepsilon+0.010\varepsilon^2)u_{xxxx}\nonumber\\
&+(-1.680-0.086\varepsilon)u
+(1.702-0.201\varepsilon-0.047\varepsilon^2)u^3\nonumber\\
&-(0.684+0.125\varepsilon)u^5, \label{highsh}
\end{align}
for $\sigma=0.01$ with a score $R^2=0.9999$ and a fit
\begin{align}
\dot{u}&=(-1.702-0.354\varepsilon+0.107\varepsilon^2)u_{xx}
+(-0.880-0.126\varepsilon+0.047\varepsilon^2)u_{xxxx}\nonumber\\
&+(-1.627-0.155\varepsilon+0.019\varepsilon^2)u
+(1.128+0.547\varepsilon-0.275\varepsilon^2)u^3\nonumber\\
&-(-0.112+1.183\varepsilon-0.357\varepsilon^2)u^5, \label{highsh2}
\end{align}
for $\sigma=0.05$ with a score $R^2=0.9977$.

\begin{figure}[h]
\includegraphics[width=0.9\linewidth]{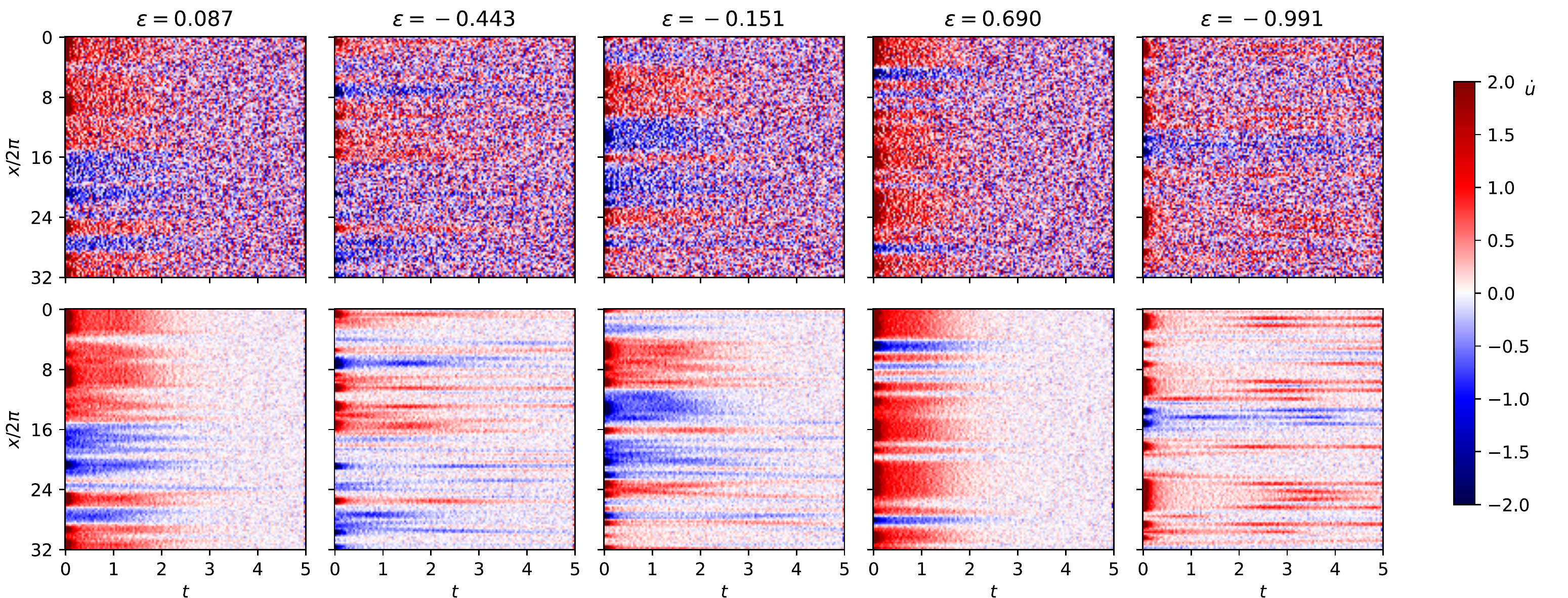}
\caption{Temporal derivatives for the low-$\varepsilon$ noisy data in the Swift-Hohenberg equation with $\sigma=0.05$ (top row) and $\sigma=0.01$ (bottom row). \label{figs6}}
\end{figure}

On the low-$\varepsilon$ training trajectories shown in Fig.~\ref{figs6}, the weak SINDyCP produces a fit
\begin{align}
\dot{u}&=(-2.001+0.004\varepsilon+0.009\varepsilon^2)u_{xx}
+(-1.002+0.005\varepsilon^2)u_{xxxx}\nonumber\\
&+(-1.692-0.075\varepsilon-0.001\varepsilon^2)u
+(1.793-0.296\varepsilon-0.021\varepsilon^2)u^3\nonumber\\
&-(0.760+0.061\varepsilon+0.008\varepsilon^2)u^5, \label{lowsh}
\end{align}
for $\sigma=0.01$ with a score $R^2=0.9999$ and a fit
\begin{align}
\dot{u}&=(-2.009+0.014\varepsilon+0.041\varepsilon^2)u_{xx}
+(-1.014-0.005\varepsilon+0.019\varepsilon^2)u_{xxxx}\nonumber\\
&+(-1.703-0.079\varepsilon+0.005\varepsilon^2)u
+(1.809-0.264\varepsilon-0.009\varepsilon^2)u^3\nonumber\\
&-(0.760+0.077\varepsilon+0.016\varepsilon^2)u^5, \label{lowsh2}
\end{align}
for $\sigma=0.05$ with a score $R^2=0.9989$.

\begin{figure}[b]
\includegraphics[width=0.9\linewidth]{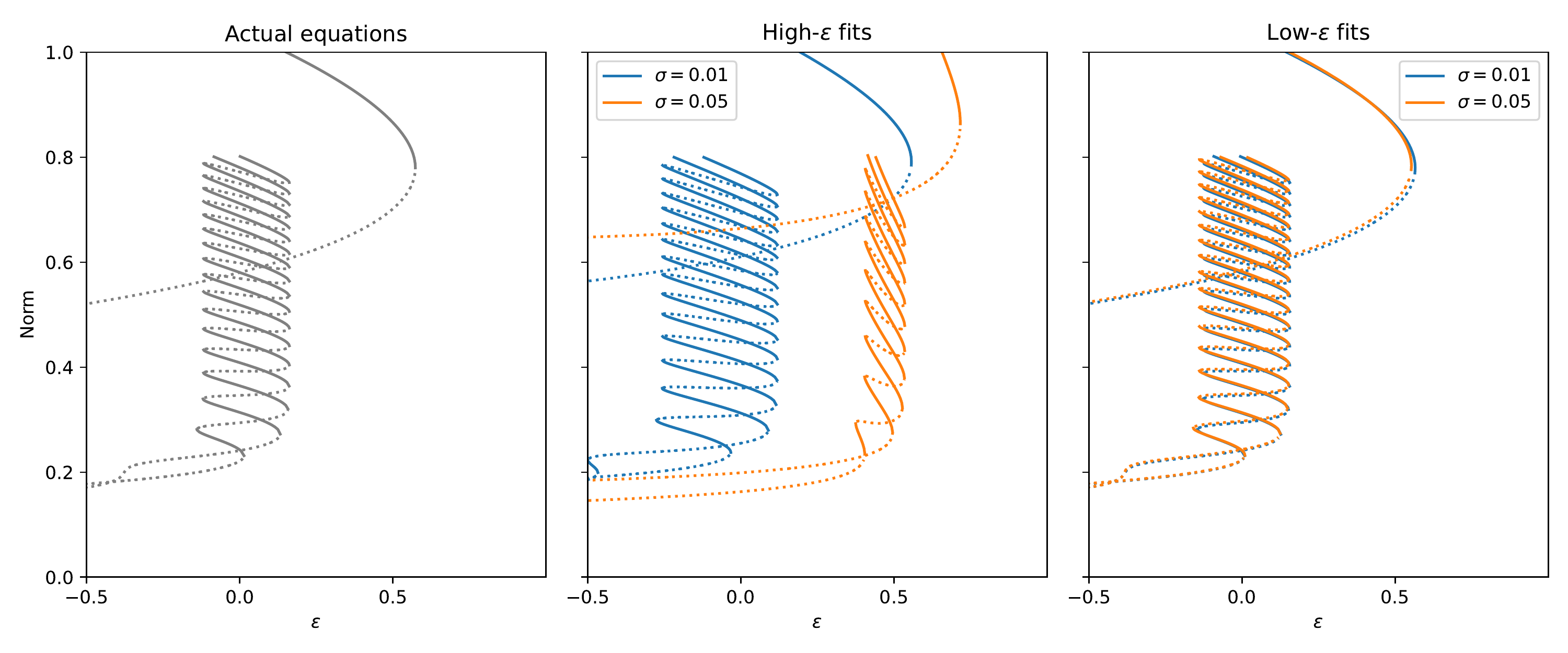}
\caption{Snaking bifurcation diagrams for the actual equations in Eq.~\eqref{originalsh} (left), the SINDyCP fit in the high-$\varepsilon$ training data in Eqs.~\eqref{highsh}-\eqref{highsh2} (middle), and the SINDyCP fit for the low-$\varepsilon$ training data in Eqs.~\eqref{lowsh}-\eqref{lowsh2} (right). \label{figs7}}
\end{figure}

We also numerically continue the localized and periodic states in AUTO for all the fits, as shown in Fig.~\ref{figs7}.
For noise intensity $\sigma=0.01$, both the high-$\varepsilon$ and low-$\varepsilon$ fits produce very good fits, but because of the greater degree of extrapolation, the details of the snaking bifurcations are slightly perturbed in the high-$\varepsilon$ case.   For noise intensity $\sigma=0.05$, the fits degrade somewhat. While the snaking bifurcations in the low-$\varepsilon$ case remain accurate, the position of the localized states becomes inaccurate in the high-$\varepsilon$ case.  Thus, it is clear that both a higher degree of extrapolation and a higher degree of noise can lead to inaccuracies when extrapolating parameters beyond the training regime.  
